\documentclass[12pt]{iopart}
\usepackage[utf8]{inputenc}
\usepackage[T1]{fontenc}
\usepackage{graphicx}
\usepackage{longtable}
\usepackage{hyperref}

\usepackage{upgreek}

\newcommand{\vect}{\mathbf}
\newcommand{\Mat}{\mathbf}
\newcommand{\text}{\mathrm}
\newcommand{\Vv}{\vect{v}}
\newcommand{\Ve}{\vect{e}}
\newcommand{\Vr}{\vect{r}}
\newcommand{\Vu}{\vect{u}}
\newcommand{\VU}{\hat{\Vu}}
\newcommand{\Vf}{\vect{f}}
\newcommand{\el}{\text{el}}

\newcommand{\etaout}{\eta_{o}}
\newcommand{\etain}{\eta_{i}}
\newcommand{\Oseen}{\mathcal{O}}

\newcommand{\deformSK}{D}
\newcommand{\shearrateSK}{\dot\gamma}

\newcommand{\inclSK}{\Psi} 
\newcommand{\phaseSK}{\phi} 
\newcommand{\phaselabSK}{\Phi} 

\newcommand{\mean}[1]{\langle {#1} \rangle}
\newcommand{\tumbSK}{\mean{\dot\inclSK}}
\newcommand{\tumbnormalized}{\omega_{\tuSCR}}
\newcommand{\visccontrastSK}{\lambda}

\newcommand{\thetacoord}{\theta} 
\newcommand{\phicoord}{\phi} 
\newcommand{\thetaref}{\thetacoord_{\text{R}}}
\newcommand{\phiref}{\phicoord_{\text{R}}}
\newcommand{\Vx}{\vect{x}}
\newcommand{\rotmat}{\Mat{D}} 
\newcommand{\rotop}{\hat{D}} 
\newcommand{\rotanglelab}{\phaselabSK} 
\newcommand{\rotangle}{\phaseSK}
\newcommand{\Vrot}{\vect{\rotanglelab}}
\newcommand{\VRot}{\hat{\Vrot}}
\newcommand{\U}{\hat{u}}
\newcommand{\Vn}{\vect{n}}
\newcommand{\transpose}[1]{#1^{T}}
\newcommand{\shape}{\beta}
\newcommand{\tuSCR}{\text{tu}}  
\newcommand{\swSCR}{\text{sw}}  
\newcommand{\trSCR}{\text{tr}}  

\begin{document}
\title{Micro-Capsules in Shear Flow}
\author{R Finken, S Kessler and U Seifert}
\address{II.~Institut für Theoretische Physik, Universit\"at
  Stuttgart, Pfaffenwaldring 57, 70550 Stuttgart, Germany}
\ead{useifert@theo2.physik.uni-stuttgart.de}
\begin{abstract}
  This paper deals with flow-induced shape transitions of elastic
  capsules. The state of the art concerning both theory and experiments is
  briefly reviewed starting with
  dynamically induced small deformation of initially spherical capsules and
  the formation of wrinkles on polymerized membranes. Initially non-spherical
  capsules show tumbling and tank-treading motion in shear flow.  Theoretical
  descriptions of the transition between these two types of motion assuming a
  fixed shape are at variance with the full capsule dynamics obtained
  numerically. To resolve the discrepancy, we expand the exact equations of
  motion for small deformations and find that shape changes play a dominant
  role. We classify the dynamical phase transitions and obtain numerical and
  analytical results for the phase boundaries as a function of viscosity
  contrast, shear and elongational flow rate.
  We conclude with perspectives on time-dependent flow, on shear-induced
  unbinding from surfaces, on the role of thermal fluctuations, and on
  applying the concepts of stochastic thermodynamics to these systems.
\end{abstract}
\submitto{\JPCM}

\section{Introduction}
\label{sec:introduction}

Elastic capsules constitute an important class within soft matter research
both from a fundamental and application point of view
\cite{pozrikidis2003book, moehwald2003, fery2007, abkarian2008,
  barthes-biesel2009}. In contrast to strongly deformable fluid vesicles
\cite{seifert1997}, the membrane of capsules exhibits finite shear
elasticity. This structural feature is achieved either by chemically or
physically crosslinking the molecules forming the membrane \cite{walter2002}
or by attaching a polymeric network to a fluid membrane as in red blood cells
(RBCs) \cite{Discher2001}. From the perspective of this paper these biological
capsules can be subsumed under this heading of a micro-capsule as
well. Promising applications of these objects arise {\it interalia} from the
pharmaceutical perspective of optimized drug encapsulation delivery and
release \cite{patwardhan1983, dinsmore2002}.

In a microfluidic context, one of the most relevant issues is the behaviour of
these soft objects in externally applied flow.
This problem is challenging from a theoretical perspective since the shape of
the capsule is not given {\it a priori} but determined dynamically from a
balance of interfacial forces with fluid stresses. Hence, their deformation
will depend on their material properties and the type of external flow.

\begin{figure}
  \centering
  \includegraphics[width=0.8\textwidth]{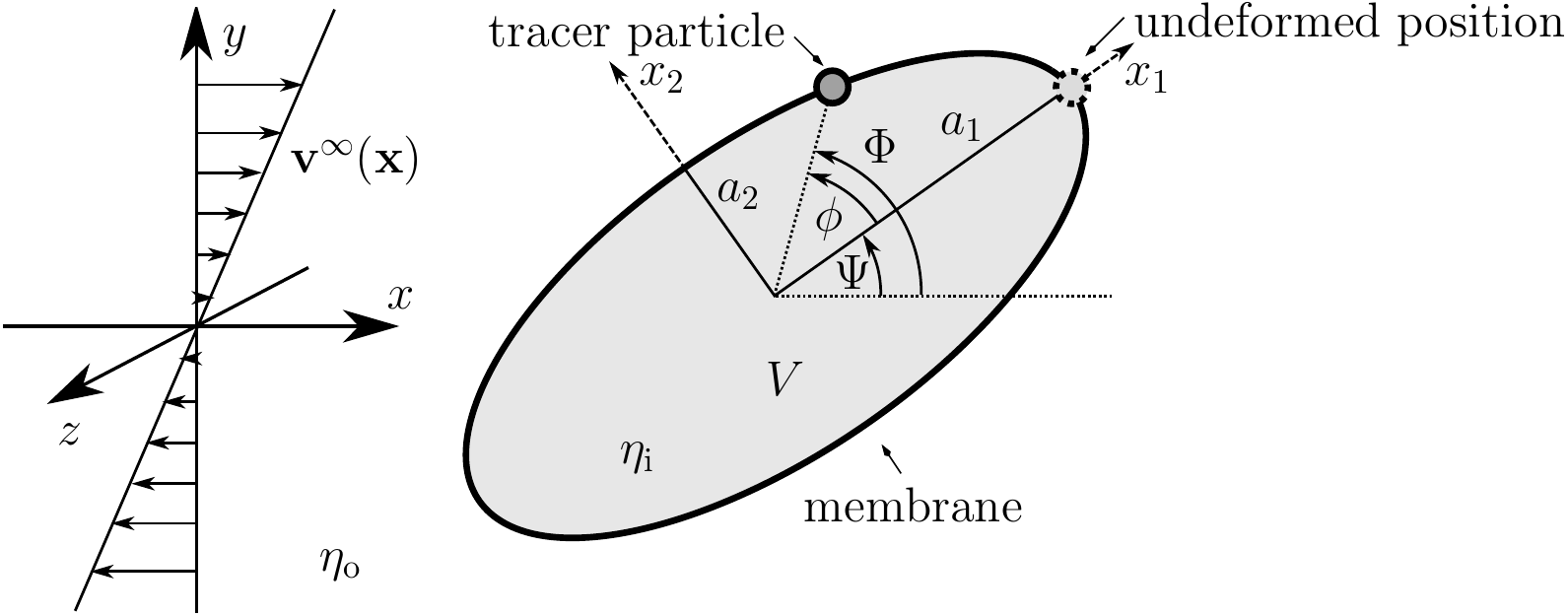}
  \caption{Ellipsoidal capsule with semi-axes $a_{1}$, $a_{2}$, $a_{3}$ in
    linear shear flow $\vect v^\infty$. The shear plane defines the $xy$ plane
    with shear flow pointing into the $x$ direction. The coordinate axes
    $x_1$, $x_2$, $x_3$ point along the principal directions of the capsule where 
    the $x_1$ and $x_2$ axes, with $a_1>a_2$, are chosen to lie in the
    shear plane and are rotated through the inclination angle $\inclSK$ with 
    respect to the $x$ and $y$ axes. 
    The phase angle $\phaselabSK$ measures the angle between the
    direction of the shear flow and the direction of a tracer particle which
    initially was on the endpoint of the long semi-axis $a_1$. The phase angle
    $\phaseSK \equiv \phaselabSK-\inclSK$ measures the tank-treading motion of the
    membrane by the angle between the position of the tracer particle with
    respect to its undeformed position.
    The membrane encloses a fluid of viscosity $\etain$ and volume $V$. The
    fluid outside the capsule has viscosity $\etaout$.}
  \label{fig:phasedef}
\end{figure}

Before reviewing the state of the art concerning theory and experiment in the
next two sections, we set the stage by introducing a few key notions for
micro-capsule dynamics. As illustrated in figure~\ref{fig:phasedef}, we
consider the dynamics of a single closed elastic membrane that surrounds an
inner fluid with viscosity $\etain$ and is embedded in an ambient flow with
viscosity $\etaout$, thereby defining an important quantity, the viscosity
contrast $\etain/\etaout$.
In the absence of the capsule we assume a prescribed external flow, in most
cases set to be an infinite linear shear flow. For all experimental setups the
Reynolds number is small, i.e.~dissipation is dominating and inertial effects
are negligible. Consequently, the flow is governed by the linear Stokes
equations and determined instantaneously by the boundary conditions, where in
all cases a no-slip boundary condition is assumed at the membrane.




For describing the capsule state adequately, we define a set of suitable
parameters. Capsules and RBCs consist of a thin membrane with a typical
thickness in the nanometer range which encloses a fluid on a larger length
scale (typically of the order 1-100 $\upmu$m) with fixed volume $V$ defining a
length scale $R$ as the radius of a sphere with the same volume
\begin{eqnarray}
  V \equiv \frac 4 3 \pi R^3 .
\end{eqnarray}
For capsules with area elasticity, the area $A$ of the membrane, as well as
the excess area $\Delta\geq0$ defined by
\begin{equation}
  A \equiv (4\pi+\Delta) R^{2} \,,
\end{equation}
can change,
where $\Delta=0$ corresponds to a sphere. In contrast, the membrane of RBCs
and polymerized vesicles has a vanishingly small area compressibility,
i.e.~$A$ and $\Delta$ are constants. For these area-incompressible objects,
shape changes are only possible for initially non-spherical shapes $\Delta>0$.

In most cases considered in this paper, the capsule is roughly ellipsoidal,
see figure~\ref{fig:phasedef} for a definition of the geometry.
The three main dynamical
parameters are (i) the inclination angle $\inclSK$ between the direction of
the shear flow and the long semi-axis, (ii) the phase angle $\phaselabSK$
measuring the angle between the direction of the shear flow and that of the
position of a tracer particle which initially was on the endpoint of the long
semi-axis and (iii) the Taylor deformation parameter
\begin{eqnarray}
  \label{eq:deform}
  \deformSK \equiv \frac{a_{1}-a_{2}}{a_{1}+a_{2}}
\end{eqnarray}
expressed by the lengths of the two semi-axes in the shear plane. For
initially spherical capsules the deformation parameter $\deformSK$ grows with
increasing shear rate, while $\deformSK$ reaches a plateau value for capsules
with incompressible membrane area at large shear rate.
\begin{figure}
  \centering
  \includegraphics[width=0.8\linewidth]{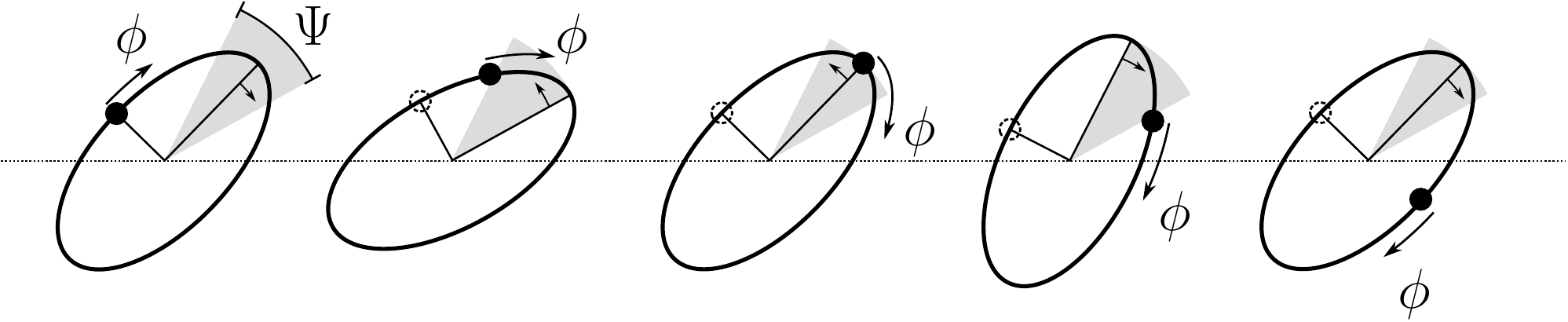}\\
  \includegraphics[width=0.8\linewidth]{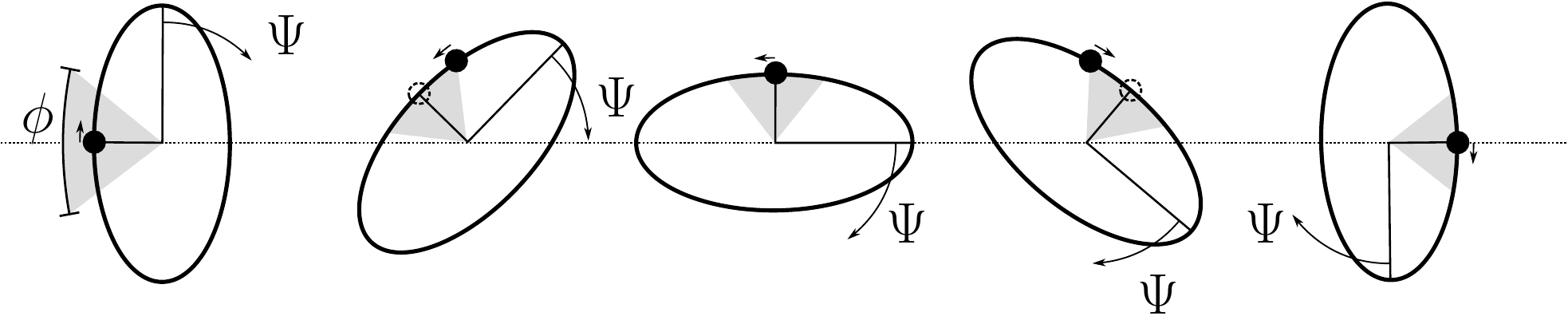}
  \caption{Sequences of swinging (top) and tumbling (bottom) of an ellipsoidal
    capsule. The black dot and arrow denote the position and motion of a
    material membrane element. In the swinging motion the inclination angle
    oscillates around a finite positive inclination angle $\inclSK$ while the
    membrane performs a tank-treading motion with monotonously decreasing
    phase angle $\phaseSK$. In the tumbling motion the inclination angle
    $\inclSK$ steadily decreases leading to a flipping motion of the capsule
    and an oscillating phase angle $\phaseSK$.}
  \label{fig:phenomenology}
\end{figure}

Two main modes of motion of capsules in shear flow can be identified as
illustrated in figure~\ref{fig:phenomenology}: (i) The capsule shape reaches a
stationary state in the laboratory frame. The membrane elements rotate along
the fixed shape like the chains of a tank. The inclination angle $\inclSK$ is
then constant, while the phase angle $\phaseSK$ grows monotonously. Such a
motion is called tank-treading in its pure form. A variant of this motion with
slightly oscillating inclination angle is called swinging and will become
important later in this paper.
(ii) The capsule shape rotates in the laboratory frame as a rigid body would
do.  In this case the inclination angle $\inclSK$ grows monotonously, while
the phase angle $\phaseSK$ stays approximately constant. This motion is called
tumbling.

This paper is organized as follows. We
give an overview over the theoretical methods in
section~\ref{sec:theor-appr}. We then summarize the experimental developments
in section~\ref{sec:experiments}.  The formation of short-wavelength wrinkles
on the surface of polymerized capsules is discussed in
section~\ref{sec:wrinkling}.  We then focus on the dynamics of non-spherical
capsules
in section \ref{sec:reducedmodel}, where we find the predictions of a simple
analytical theory at odds with numerical evidence. To solve this discrepancy
in the main original contribution of this paper, we systematically expand the
equations of motion in the limit of small capsule deformations in
section~\ref{sec:syst-quasi-spher}. We find that in certain dynamic regimes
shape changes, which are not covered by the simple theory, play an important
role in the capsule dynamics.
We finish by discussing a few perspectives
in section~\ref{sec:perspectives}. Details of the analytical methods and
longer derivations are deferred to
appendices~A--C.

\section{Theoretical approaches}
\label{sec:theor-appr}
From a theoretical point of view, the full equations of motion are intractable
analytically. However, much insight into the physics can be obtained by either
studying certain asymptotic limits, where the dynamics can be simplified, or
by restricting the number of degrees of freedom to as few as possible. An
example of the first strategy is the quasi-spherical expansion of the
equations of motion for initially spherical capsules as shown in the following
section~\ref{sec:init-spher-caps}. The Keller-Skalak model \cite{keller1982}
(described in detail in section \ref{sec:keller-skalak}) follows the second
strategy, by artificially fixing the shape of the capsule and only allowing
tank-treading and tumbling degrees of freedom. In its original form this model
is only applicable for fluid vesicles although it was intended to describe
RBCs. It is missing an important energetic contribution to the dynamics, the
so-called shape memory \cite{fischer2004}. The origin of this term is
discussed in section~\ref{sec:shape-memory-effect}. An extension of the
Keller-Skalak model including the shape memory effect \cite{skotheim2007,
  abkarian2007} is described in detail in
section~\ref{sec:skotheim-secomb}. Beyond these two approaches, numerical
methods are inevitable. In section \ref{sec:numerics}, we give an overview of
the different numerical approaches that have been developed and successfully
used to determine the dynamics of vesicles, capsules and RBCs.

\subsection{Initially spherical capsules}
\label{sec:init-spher-caps}


Often the shape of the capsule deviates only little from a sphere. Similarly,
in weak flow, initially spherical capsules only deviate slightly from their
equilibrium shape. The highly non-linear equations of motion can then be
expanded in terms of the small deformation vector field. One can then take
further advantage of the high symmetry of a sphere by changing to a spherical
harmonic representation. A comprehensive treatment is illustrated in
\cite{rochal2005}, where the displacement field is expanded in
vector spherical harmonics, which are defined for completeness in
\ref{sec:vect-spher-harm} since we use them in our approach as well.






In early work, Barth{\`e}s-Biesel \cite{barthes-biesel1980} considered the
steady state of an initially spherical elastic capsule using a small
deformation expansion. The membrane is assumed to be homogeneous, isotropic,
area-incompressible, and shows no resistance against bending. To linear order
in the deformation only tank-treading motions are possible in the long time
limit.  Only at higher order a transition to tumbling can occur, when the
shear forces have deformed the capsule sufficiently. In the steady state, the
capsule is orientated with inclination angle $\inclSK=\pi/4$ while the
membrane performs a tank-treading motion. With increasing shear rate, the
deformation first increases linearly with quadratic corrections. In
\cite{barthes-biesel1981}, the relaxational dynamics of a capsule with
arbitrary elastic response was obtained still excluding bending energy. There,
results have also been obtained in limit cases, like droplets (showing only
surface tension) or red blood cells (having a local area incompressibility).
More recent theoretical efforts are reviewed by D.~Barth{\`e}s-Biesel and
C.~Pozrikidis in Chapter 1 and 2, respectively, of \cite{pozrikidis2003book}.
Bending energy was first included for capsules by Rochal \etal
\cite{rochal2005}.

\subsection{Shape memory effect: non-spherical capsules}
\label{sec:shape-memory-effect}

For a capsule with a spherical equilibrium shape all membrane points are
equivalent from a mechanical point of view. This is no longer true for an
initially non-spherical capsule such as an RBC, since the long and short axes
are distinguished points on the membrane. For a discussion of the consequences
for the energetics of a tank-treading motion let us assume that the capsule
membrane is unstressed in the equilibrium state. As with tank-treading motion,
we now rotate the membrane by the phase angle $\rotangle$. Since the local
geometry of the capsule of each membrane element is different from that of the
starting position, the capsule membrane will be sheared, requiring a positive
elastic energy. Consequently, a force aims to restore the unstressed
equilibrium state. Only after a rotation by $\rotangle=\pi$ the initial and
final position of each membrane point are equivalent to each other due to the
mirror symmetry of the ellipsoidal capsule. The membrane is then again
unstressed.  Any membrane element therefore energetically prefers its initial
position (or one of the equivalent positions by symmetry of the reference
shape). This property is called shape memory effect \cite{fischer2004}, and
plays an important role also in the dynamics of red blood cells
\cite{takatani2006, fischer2007, abkarian2007, skotheim2007}.


A direct consequence of the shape memory effect is that capsules perform
tumbling motions for low shear rates. Here, the hydrodynamic forces are not
sufficient to overcome the elastic barrier which is present for a hypothetical
tank-treading motion.  Only for higher shear rates the elastic barrier can be
overcome and tank-treading motion becomes possible. Since the force balance of
a capsule is also slightly changed by the shape memory effect, one now
observes stable oscillations of the inclination angle $\inclSK$ around a
finite value instead of a stationary tank-treading state. This type of motion
has therefore been called swinging motion \cite{abkarian2007}. This term
should not be confused with the swinging mode of fluid vesicles, which is also
called vacillating-breathing or trembling \cite{misbah2006, noguchi2007,
  lebedev2007a}.

\subsection{Full numerical treatment}
\label{sec:numerics}


For quantitative results beyond asymptotic limits one has to resort to
numerical methods for capsules in shear flow. Boundary integral methods (BIM)
have been used as first approaches in handling capsules in flow \cite{li1988,
  leyrat-maurin1993, leyrat-maurin1994, zhou95, pozrikidis1995, ramanujan1998,
  pozr01, pozrikidis2003a, lac03, lac2004, lac2005}. As a mixture of global
spectral methods \cite{kessler2007} and boundary integral methods, the
spectral collocation \cite{pozrikidis2006a} and the spectral boundary methods
\cite{Wang2006, dimitrakopoulos2007, dodson2008} have been developed which use
higher order basis functions on a triangulated surface.

The immersed boundary method \cite{peskin1977, peskin2002} can be implemented
in any hydrodynamic solver. Here, the force density of the boundary is mapped
onto a Cartesian mesh. It has been combined with solvers of the Navier-Stokes
equation \cite{eggleton1998, bagchi2009} and with lattice Boltzmann methods
\cite{sui2007, sui2007a, sui2007b, sui2008, sui2008a, sui2010, sui2010a}. The
front tracking technique \cite{unverdi1992} can handle two-phase flows and was
used with a finite volume method to treat two-dimensional capsules
\cite{ma2009}.

The oscillations of the shape and orientation of non-spherical capsules have
first been observed by Ramanujan \etal \cite{ramanujan1998} who used a
boundary spectral method. For high and low deformations this method was
plagued by numerical instabilities due to the degradation of the grid. Further
improvement of the boundary element method allows the stable simulation of
swinging and tumbling of highly flattened capsulses only by
numerically smoothing the surface \cite{pozrikidis2003a}. More recent
approaches, such as the spectral boundary algorithms
\cite{Wang2006, dimitrakopoulos2007, dodson2008} use higher order basis
functions on the triangulated surface to avoid numerical instabilities.  Pure
spectral methods \cite{kessler2007} employ global basis functions to
parameterize the capsule membrane. The fluid dynamics part is also simulated
using lattice Boltzmann methods
\cite{low2007, sui2008, sui2010, kaoui2008}. Recently, front-tracking methods
have been employed to study the motion of two-dimensional \cite{ma2009} or
three-dimensional capsules \cite{bagchi2009, sui2010a}.

Existing solvers for the dynamics of vesicles, which treat the flow at a
continuum level, either employ the boundary integral method on a discrete
triangulation scheme \cite{kraus1996, kaoui2008, biben2009} or a phase field
or advected field model \cite{biben2003a, beaucourt2004b, biben2005}. An
alternative route is provided by stochastic rotation dynamics (SRD) or
multi-particle collison dynamics (MPCD), where the flow is modeled using
effective fluid particles interacting with a dynamically triangulated membrane
\cite{noguchi2004, noguchi2005, noguchi2005a, noguchi2007, mcwhirter2009}.

\section{Experiments}
\label{sec:experiments}


Early experiments \cite{chang1993b, chang1993, walter2000, walter2001}
confirmed the quasi-spherical predictions for the deformation parameter
$\deformSK$ (for not too high shear rates) as well as the decreasing behaviour
of the inclination angle $\inclSK$ for increasing shear rate.  Although the
inclination angle $\inclSK$ deviated from the theoretical prediction $\pi/4$
even in the limit of vanishing shear rates $\shearrateSK=0$ \cite{chang1993,
  walter2001}, experiments together with the quasi-spherical results were used
to infer values for the elastic parameters of the membrane.

In contrast to the theoretical predictions for spherical shapes, capsules in
such experiments \cite{chang1993, walter2001} showed oscillations of both the
deformation $\deformSK$ and the inclination angle $\inclSK$. These
oscillations caused by the fact that real capsules deviate from a perfectly
spherical shape are a direct consequence of the shape memory effect discussed
above in section~\ref{sec:shape-memory-effect}.


In order to observe tumbling motion for experimentally accessible shear rates,
capsules which show a large deviation form the spherical shape have to be
used. Red blood cells are an ideal and convenient choice. In early
experiments, red blood cells have only be examined either in a tumbling state
in the blood plasm or in a swinging state for high shear rates showing a
stationary inclination angle \cite{goldsmith1972, fischer1977, fischer1978,
  transontay1984}. The motion for low shear rates and close to the
swinging transition was experimentally examined by Abkarian \etal
\cite{abkarian2007}.  Within the swinging regime the frequency of the
oscillation of the inclination angle is exactly twice the frequency
corresponding to the tank-treading motion. While a membrane point is advected
a full rotation along the membrane, the inclination angle is performing two
full swinging motions. This relation highlights the direct connection between
the oscillations of the shape and the motion of the membrane. For a decreasing
shear rate within the swinging regime, the frequency of the oscillation
decreases, while the oscillation amplitute of inclination angle increases. The
numerical analysis of the full equations of motion using spectral methods
\cite{kessler2007} discussed in section~\ref{sec:numerics} yields qualitative
and quantitative agreement with the experimental results. Interestingly, a
hysteresis effect at the transition between tumbling and swinging occurs when
alternatingly increasing and decreasing the shear rate
\cite{abkarian2007}. Here, a single tumbling motion seems to mix with
multiple swinging motions.

\section{Wrinkling}
\label{sec:wrinkling}

\begin{figure}
\centering
\includegraphics[width=0.3\linewidth]{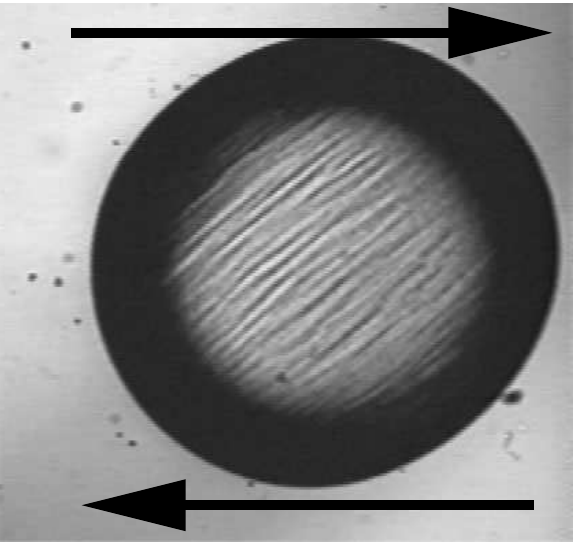}   
\caption{Wrinkling of a polysiloxane capsule ($\simeq 343 \mu {\rm m}$) in
  shear flow as indicated by the arrows (adapted from \cite{walter2001}).}
\label{fig:wrinkling}
\end{figure}

Rather than deforming to a slightly ellipsoidal shape, certain
polymeric capsules show a wrinkling instability while
tank-treading in a linear shear flow \cite{walter2001}, see
figure~\ref{fig:wrinkling}. The wrinkles are caused by a
compression-force acting onto the membrane. Although the membrane
moves along the stationary shape while tank-treading, both
position and direction of the wrinkels remain constant as they
are caused by the compression part of the elongational part of
the flow. Thin membranes are unstable with respect to compressive
forces. In a flat geometry, compression leads to Euler buckling
on the largest possible length scale. To produce small length
scale wrinkles on a flat membrane one needs a combination of
tensile stress along the wrinkles and a non-local geometric
constraint \cite{cerda2003}.

\begin{figure}
  \centering
      \includegraphics[width=\linewidth]{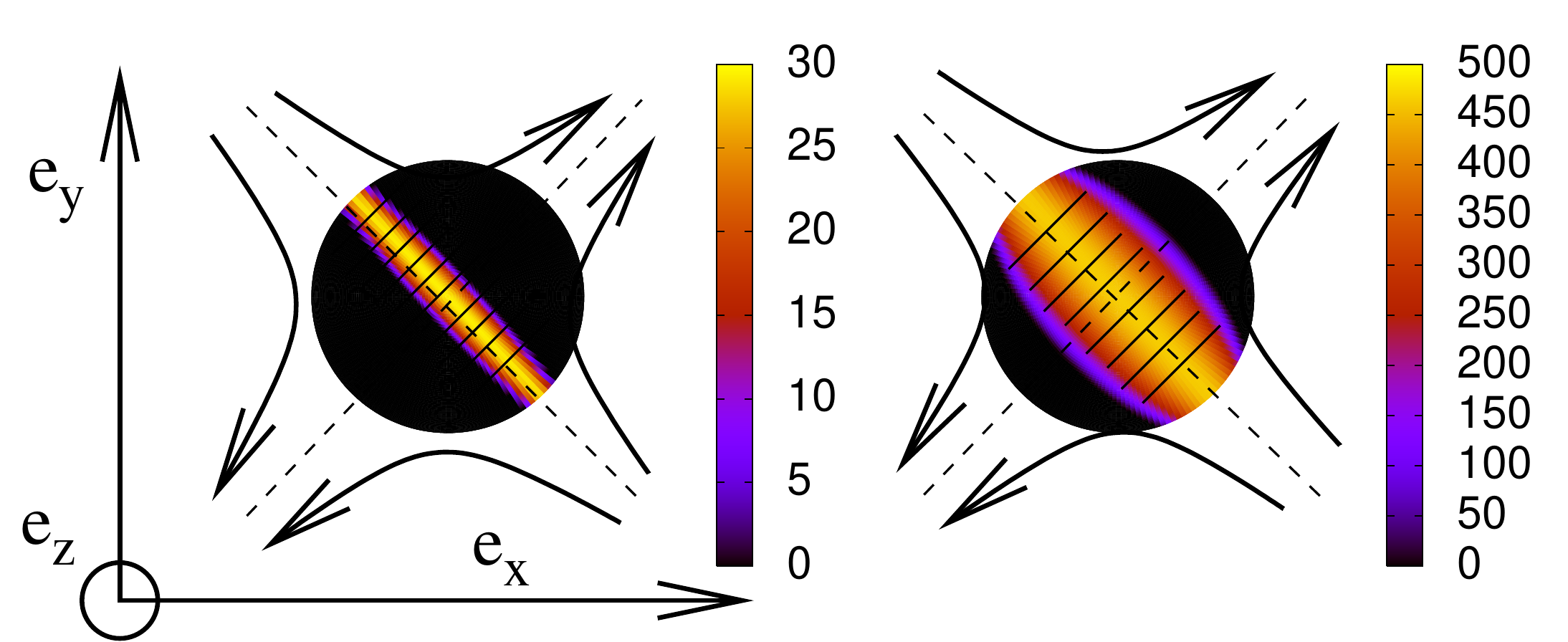}
      \caption{Wrinkling instability region for $\dot{\gamma}=11.5 {\rm
          s^{-1}}$ (left) and $\dot{\gamma}=17.0 {\rm s^{-1}}$ (right),
        respectively.  The elastic moduli of the membrane are $K = 0.2
        \mathrm{N m^{-1}}$, $\kappa = 1\cdot 10^{-17} \mathrm{Nm}$. In the
        black coloured region the membrane is still stable with respect to
        wrinkle formations. Brighter colons indicate a larger growth rate of
        the wrinkles. Initially, wrinkling occurs only in a small region
        around the circumference corresponding to the plane of maximal
        compressive stress.  At higher shear rates, this region grows, as does
        the growth rate of each wrinkling mode. The direction of the wrinkles
        along the lines of maximal tension are indicated schematically
        (adapted from \cite{finken2006}).}
   \label{fig:region}
\end{figure}

In the small deformation regime this wrinkling instability was analytically
explained in 
\cite{finken2006} using a separation of length scales,
since the capsule radius is much larger than the typical wavelength of the
wrinkles. Even though the overall dynamics of a capsule in viscous flow is
inherently non-Hamiltonian, energy considerations dominate the physics at the
small length scale of the wrinkles. The energy required to form wrinkles of
amplitude $\delta{r}$ and local wave number $k$ is
\begin{equation}
  \mathcal{H}^{\text{el}} \sim \mathcal{H}^{\text{el}}_{0} + \int dA 
  K (\delta r)^{2}/R^{2} + \sigma_{0} (\delta r)^{2} k^{2}  + \kappa
  (\delta r)^{2}  k^{4}.
\end{equation}
Here, $K$ is the area incompressibility and $\sigma_{0}<0$ the
compression produced by the hydrodynamic force. This term drives
the instability. Large wave numbers $k$ are penalized
energetically by the bending energy $\sim \kappa k^{4}$. In
contrast to a flat membrane, buckling on any length scale on a
curved membrane costs elastic energy $\sim K/R^{2}$, since the
area of the membrane is increased (decreased) when a membrane
element is moved outwards (inwards). The resulting tensile
(compressive) stress suppresses buckling on too large scales.
Only when the compressive force is large enough formation of
wrinkles can reduce the elastic energy in the system. This
happens at intermediate wave numbers, which can be calculated by
minimizing above energy.

For polymerized capsules, the analytical expression for the critical shear
rate above which the wrinkling starts reads
  \begin{equation}
    \dot{\gamma}_{c} = \frac{8}{5 \eta_{\text{o}} R^{2}}
    \sqrt{\mu \kappa \frac{K}{K + \mu}},
  \end{equation}
and the critical wavelength of the wrinkles becomes
\begin{equation}
    k_{c} = \sqrt[4]{\frac{4 \mu}{R^{2}\kappa} \frac{K}{K +
        \mu}}.
\end{equation}  
At $\dot{\gamma}_{c}$, the wrinkles start to form along a circumference which
forms an angle of $\pi/4$ with respect to the shear direction. At higher shear
rates $\dot{\gamma} > \dot{\gamma}_{c}$, wrinkling extends over larger regions
on the capsule, as shown in figure~\ref{fig:region}. The analytical expression
for the critical shear rate would allow for an experimental determination of
the bending rigidity of the membrane which is difficult to obtain
otherwise. So far, experimental work seems not to have taken up this
suggestion.

\section{Reduced model with two degrees of freedom}
\label{sec:reducedmodel}

We now turn to capsules which are initially non-spherical.

\subsection{Keller-Skalak model}
\label{sec:keller-skalak}

In the theory of Keller and Skalak \cite{keller1982}, a red blood cell (RBC)
is treated within a simplified model. Since as an essential ingredient the
shape memory effect of RBCs is missing within this simplified description, the
model does not capture essential parts of the physics. Yet, for vesicles the
Keller-Skalak model delivers quite a good description. Since it is also the
basis for the further modelling of capsules,
we briefly recall its main features.
The three-dimensional RBC or vesicle is assumed to have a fixed ellipsoidal
shape
\begin{equation}
  (x_1/a_1)^2 + (x_2/a_2)^2 + (x_3/a_3)^2 = 1,
\end{equation}
see figure~\ref{fig:phasedef}.

The (Cartesian) components of the undisturbed shear flow are $(\shearrateSK
y,0,0)$. In order to mimic a tank-treading motion of the membrane, the
velocity field at the membrane is assumed to have components
\begin{equation}
  \Vv  = \omega^{\text{KS}}_{\text{tt}} 
  \left( -(a_1/a_2)x_2,(a_2/a_1)x_1,0 \right),
\end{equation}
where $\omega^{\text{KS}}_{\text{tt}}$ is a parameter having the dimensions of
a frequency. The energy supplied by the external flow has to be balanced with
the energy dissipated inside the RBC. Likewise the torques acting on the RBC
have to balance. The equation of motion of the RBC derived from this torque
balance reads \cite{keller1982}
\begin{equation}
  \dot{\inclSK} 
  = -\left(\frac{\dot{\gamma}}{2}
    +\frac{2a_{1}a_{2}}{a_{1}^{2}
      +a_{2}^{2}}
    \dot\phaseSK \right) 
  + \frac{a_{1}^{2}-a_{2}^{2}}{a_{1}^{2}+a_{2}^{2}} 
  \frac{\shearrateSK}{2}  \cos(2 \inclSK) \,.
  \label{eq:kstorque}
\end{equation}
The dissipated energy must come from the work performed by the external flow,
which leads to the equation
\begin{equation}
  \label{eq:ksenergy}
  \etain f_1 \dot{\phaseSK}^{2} = 
  \etaout (f_2\dot{\phaseSK}^{2} 
  + f_3 \shearrateSK \dot{\phaseSK} \cos(2\inclSK)) \,,
\end{equation}
where the geometric quantities $f_i$ depend only upon the ratios
of the axis lengths
$a_i$ and are defined by
\begin{eqnarray}
  \eqalign{
  \alpha_i \equiv a_i (a_1a_2a_3)^{-1/3}~\mathrm{for}~i=1,2,3 \,,~
  z_1 \equiv \frac 1 2 \left(\frac{a_1}{a_2} - \frac{a_2}{a_1}\right) \,,\\
  z_2 \equiv (\alpha_1^2+\alpha_2^2)\int_0^\infty 
  (\alpha_1^2+s)^{-3/2}(\alpha_2^2+s)^{-3/2}(\alpha_3^2+s)^{-1/2}
  \mathrm{d}s \,,\\
  f_1 \equiv 2z_1^2 \,,~
  f_2 \equiv 4z_1^2\left(1-\frac{2}{z_2}\right) \,,~
  f_3 \equiv -4 \frac{z_1}{z_2} \,.
  }
\end{eqnarray}
Both balance laws can be simplified to
\begin{eqnarray}
  \label{eq:kseom}
  \dot \inclSK &=& \shearrateSK(A + B \cos(2 \inclSK)) \,, \\
  \dot \phaseSK &=& \shearrateSK C \cos(2\inclSK)
\end{eqnarray}
with
\begin{eqnarray}
  A &\equiv& - 1/2 \,, \\
  B &\equiv& \frac{
    a_{1} a_{2} f_{3} + (f_{2}-f_{1}\etain/\etaout)(a_{1}^{2}-a_{2}^{2})}
  {2(a_{1}^{2}+a_{2}^{2})(f_{2}-f_{1}\etain/\etaout)}
  \,, \\
  C &\equiv& - \frac{f_{3}}{f_{2}-f_{1}\etain/\etaout} \,.
\end{eqnarray}

The phase diagram for the dynamics of RBCs or vesicles is rather simple in the
Keller-Skalak model. The transition between tumbling and tank-treading is
independent of the shear rate. For given half axes $a_i$, there is a critical
viscosity contrast $(\etain/\etaout)_{\text{c}}$ given by $|B|=|A|=1/2$, below
which the RBC or vesicle performs a tank-treading motion. For larger viscosity
contrast, it undergoes a tumbling motion. The transition from tumbling to
tank-treading is thus independent of the shear rate $\shearrateSK$ within the
Keller-Skalak model.

For low viscosity contrast $\etain/\etaout < (\etain/\etaout)_{\text{c}}$,
i.e.~$|B|>1/2$, one obtains a steady tank-treading angle
\begin{equation}
  \label{eq:46}
  \inclSK_{\text{tt}}
  = \frac{1}{2} \arccos\left(\frac{1}{2B}\right)
\end{equation}
and a steady tank-treading frequency
\begin{equation}
  \omega^{\text{KS}}_{\text{tt}}
  = \dot{\gamma} \frac{C}{2B} \,.
  \label{eq:ksomegatt}
\end{equation}
For high viscosity contrast $\etain/\etaout > (\etain/\etaout)_{\text{c}}$,
i.e.~$|B|<1/2$, one obtains a steady tumbling motion with mean tumbling rate
\begin{equation}
  \tumbSK  \equiv
  \lim_{T\to\infty} \frac{1}{T} \int_0^T \dot\inclSK(t') dt'
  = -\frac{\shearrateSK}{2} \sqrt{1-4B^2}.
\end{equation}

\subsection{Skotheim-Secomb model}
\label{sec:skotheim-secomb}

The Keller-Skalak model in its original form can only correctly
describe the motion of vesicles, as it does not take into account
the shape memory effect. Abkarian \etal \cite{abkarian2007} and
Skotheim \etal \cite{skotheim2007} extended the Keller-Skalak
model to remedy this deficiency.

The membrane with its elastic property serves as an energy storage and induces
an additional torque onto the capsule. The elastic energy
$E(\rotangle)$ required for a hypothetical tank-treading motion
must be positive and periodic in the phase angle $\phaseSK$ with
period $\pi$, as argued in section~\ref{sec:shape-memory-effect}.
In 
\cite{abkarian2007, skotheim2007},  the energy was
modeled as a simple harmonic function
\begin{eqnarray}
  E(\phaseSK) = E_0 \sin^2{\phaseSK} \,.
\end{eqnarray}
A Keller-Skalak derivation balancing energy and torque leads to two equations
of motion for the two degrees of freedom, the inclination angle and the phase
angle.
\begin{eqnarray}
  \label{eq:skotheimeoma}
  \dot \inclSK(t) &=& \tilde A(t) + \tilde B \cos{(2\inclSK(t))}
  , \\
  \dot \phaseSK(t) &=& \tilde C \left(
    \cos{(2\inclSK(t))}
    - \frac{E_0}{V \etaout f_3 \shearrateSK} \sin{(2\phaseSK(t))}
  \right)
  \label{eq:skotheimeomb}
\end{eqnarray}
with constants
\begin{eqnarray}
  \tilde A(t) &\equiv& -\left( \frac 1 2 \shearrateSK
    + \frac{2a_1 a_2}{a_1^2+a_2^2}
    \dot \phaseSK(t) \right),\\
  \tilde B &\equiv&
  \frac{a_1^2-a_2^2}{2(a_1^2+a_2^2)}\shearrateSK,\\
  \tilde C &\equiv& -\frac{f_3}{f_2-f_1 \etain/\etaout} \shearrateSK .
\end{eqnarray}

In dimensionless quantities of shear rate $\chi$, viscosity ratio $\lambda$
and eccentricity $\alpha$
\begin{eqnarray}
  \chi &\equiv& \frac{V \etaout (-f_3)}{E_0} \shearrateSK,\\
  \lambda &\equiv& \frac{f_1}{-2f_3} \frac{\etain}{\etaout} + \frac{-f_2}{-2f_3},\\
  \alpha &\equiv& \arctan \frac{a_1^2-a_2^2}{2a_1a_2},
\end{eqnarray}
the equations of motion can be cast in simple form \cite{kessler2009}
\begin{eqnarray}
  \dot\inclSK(\tau)
  &=& -\cos\alpha \, \dot \phaseSK(\tau)
  -\lambda (1 - \sin\alpha \, \cos 2\inclSK(\tau) ) \,,
  \label{eq:dim1} \\
  \dot \phaseSK(\tau)
  &=& - (\chi^{-1}(\tau) \sin2\phaseSK(\tau)
  + \cos 2\inclSK(\tau) ) \,.
  \label{eq:dim2}
\end{eqnarray}
Here, where $\tau$ is a dimensionless time defined differentially as
\begin{equation}
 d\tau \equiv \frac{2\shearrateSK}{\lambda} dt.
\end{equation}

Solving the equations of motion within the reduced model for
arbitrary parameters has been performed numerically
\cite{abkarian2007, skotheim2007, noguchi2009a}. Kessler \etal
\cite{kessler2009} have obtained analytical solutions in the
quasi-spherical limit and, using of the method of asymptotic
expansion \cite{hinch1991}, both the stable trajectories and the
slope of the transition between tumbling and swinging.

For small shear rates $\shearrateSK$ the capsule always tumbles as the
hydrodynamic forces are too weak to overcome the elastic barrier. For
increasing shear rate the transition to the intermittent motion takes
place. By further increasing the shear rate, the transition to swinging occurs
for small and moderate viscosity contrast $\visccontrastSK$.

\begin{figure}
  \centering
  \includegraphics[width=0.48\linewidth]{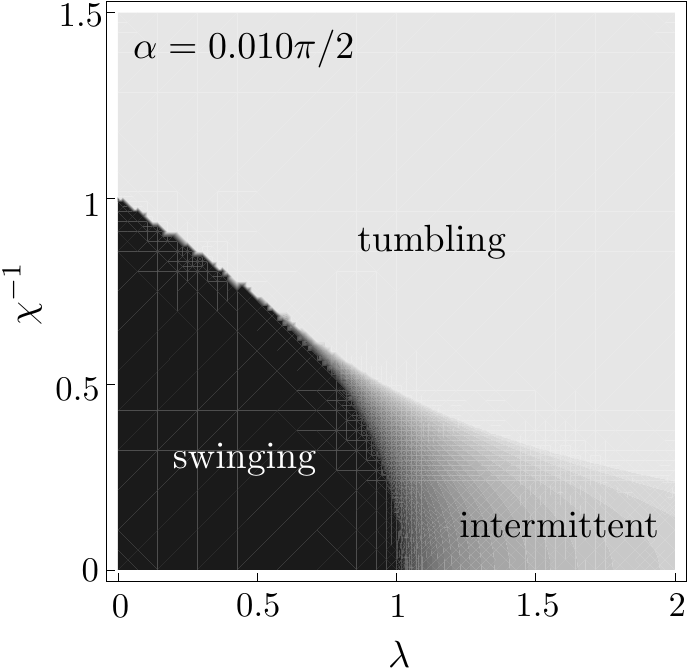}
  \includegraphics[width=0.48\linewidth]{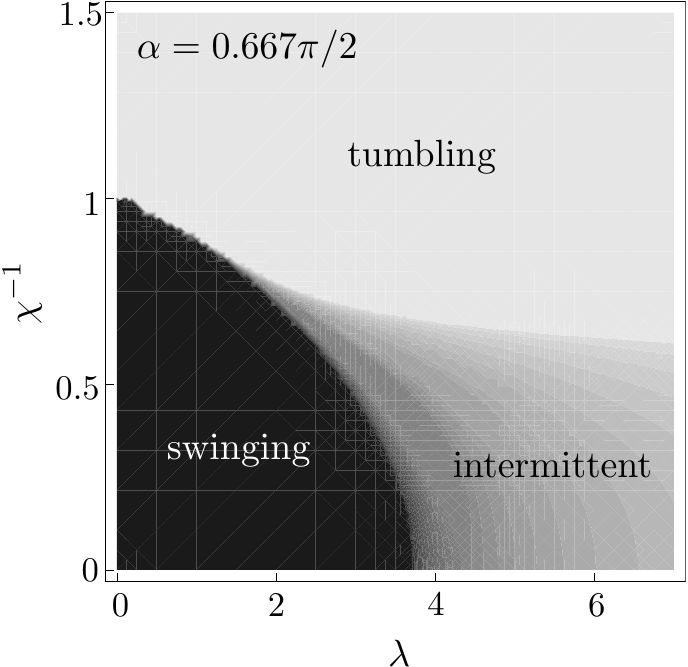}
  \caption{Grey scale plots of the normalized mean tumbling rate $\tumbSK$
    within the reduced model for two different ellipsoidal reference shapes
    with eccentricity $\alpha$, $\alpha=0$ corresponding to a spherical shape
    within the shear plane. Dark grey colour corresponds to a vanishing mean
    tumbling rate $\tumbnormalized=0$, while light grey corresponds to pure
    tumbling $\tumbnormalized=1$.}
  \label{fig:reducedphasediagram}  
\end{figure}

The phase diagram (see figure~\ref{fig:reducedphasediagram}) of this reduced
model
shows the regimes of the three possible motions. Either the capsule tumbles,
swings or performs an intermittent motion with mixed tumbling and swinging
periods. The normalized mean tumbling rate
\begin{eqnarray}
  \tumbnormalized \equiv \frac{\mean{\dot \inclSK}}
  {\mean{\dot \inclSK}+\mean{\dot \phaseSK}}
  \equiv \lim_{T\to\infty}
  \frac{\int_0^T \dot \inclSK(t) dt/T}{\int_0^T \dot \inclSK(t) dt/T
  + \int_0^T \dot \phaseSK(t) dt/T}
\end{eqnarray}
serves as an order parameter to distinguish between the possible motions
\cite{kessler2009}. Pure tumbling motion is characterized by
$\tumbnormalized=1$, pure swinging by $\tumbnormalized=0$, while for
intermittent motion the mean tumbling rate $\tumbnormalized$ takes values
between 0 and 1. Noguchi has pointed out that $\tumbnormalized$ reaches
several plateau values between 0 and 1 in the intermittent regime
\cite{noguchi2009a}.

\subsection{Comparison with spectral method}
\label{sec:comp-with-numer}

To check the validity of the reduced model, its predictions have
to be compared with the full non-linear dynamical motion obtained by
numerical methods.  The first systematic numerical phase
diagram for a large range of shear rate and viscosity contrast
was constructed using the spectral method described in
section~\ref{sec:numerics} \cite{kessler2007}. Using this method,
tumbling motions for capsules slightly deviating from the
spherical shape were observed numerically for the first time.

\begin{figure}
  \centering
  \includegraphics[width=0.9\linewidth]{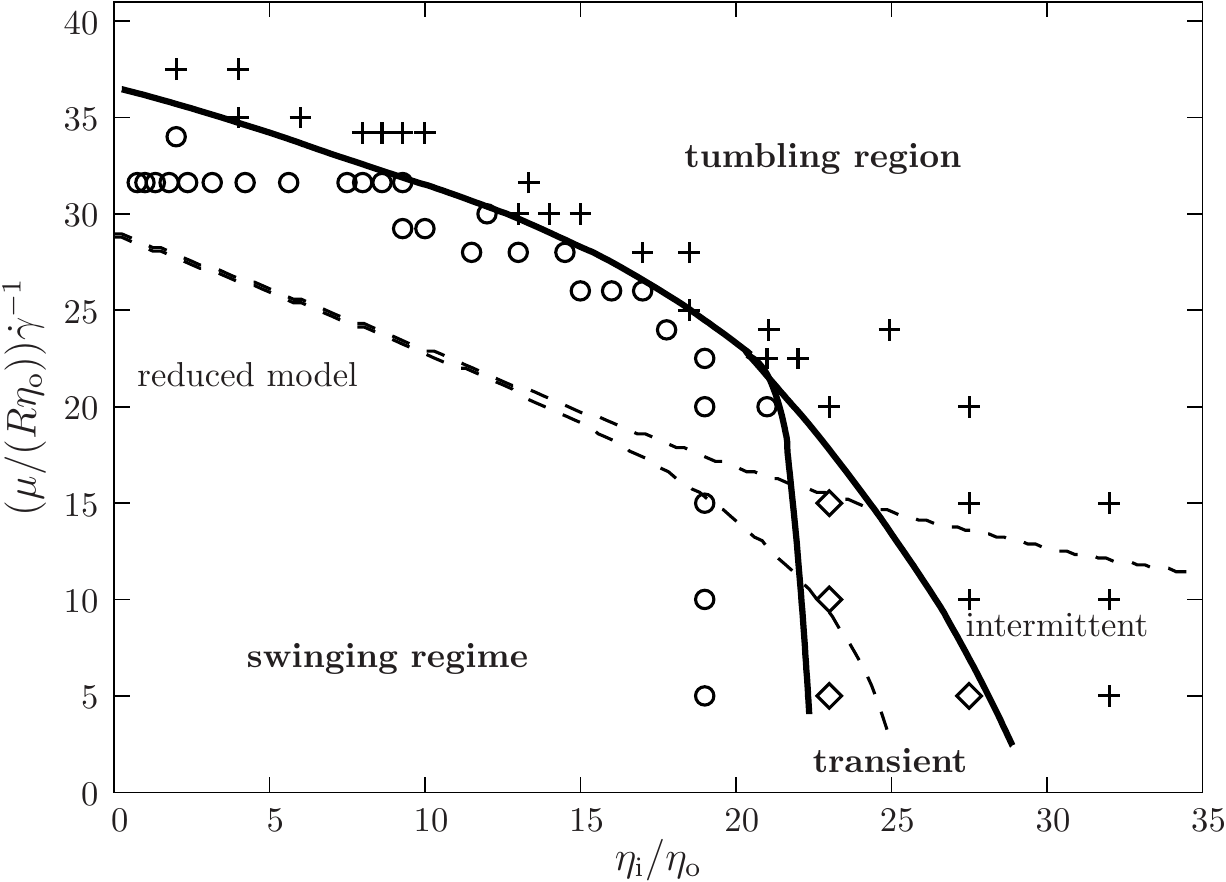}
  \caption{Comparison of phase diagrams deduced by the spectral method
    \cite{kessler2007} (symbols and full line) with the one following from the
    reduced model (dashed lines). Phase diagrams are functions of the
    viscosity contrast $\etaout/\etain$ and the inverse shear rate
    $(R\etaout \shearrateSK/\mu)^{-1}$, where $\mu$ is the shear
    modulus of the capsule. For low shear rates the capsule
    tumbles while for higher shear rates the capsule performs swinging motions
    for a not too high viscosity contrast. In between these two regimes, there
    is either a transient regime or intermittent regime deduced by the
    spectral method or the reduced model, respectively. Adapted
    from \cite{kessler2007}.}
  \label{fig:phasediagramkessler}
\end{figure}

The phase diagram obtained from the spectral method 
as shown in figure~\ref{fig:phasediagramkessler} shows qualitative agreement
with the predictions of the reduced model. The reduced model correctly
predicts a swinging and tumbling motion and a transition for increasing shear
rate. In particular for small viscosity ratios, the slopes of the
tumbling-swinging transition agree quantitatively, while the absolute position
of the transition differs by about $20$ percent.

\begin{figure}[t]
  \centering
  \includegraphics[width=0.4\linewidth]{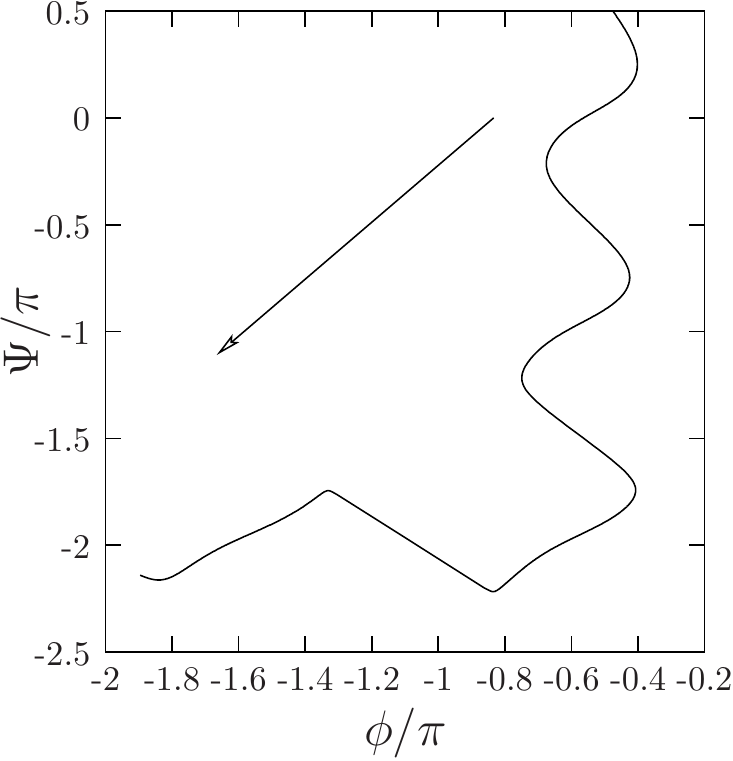}
  \caption{Transient motion from tumbling to stable swinging motion. In this
    parametric plot the transient trajectory is shown as a function of the
    inclination angle $\inclSK$ and the phase angle $\phaseSK$ where the arrow
    denotes the direction of time. Initially, the capsule tumbles with
    oscillating $\phaseSK$ and steadily decreasing $\inclSK$, but finally ends
    up in a swinging motion, where $\inclSK$ begins to oscillate around a
    stable value. Adapted from \cite{kessler2007}.}
  \label{fig:transient}
\end{figure}

Yet, in the intermittent regime of the reduced model, only transient motion
was found in the full model as can be seen in figure~\ref{fig:transient}.
Here, the capsule initially tumbles with an steadily decreasing inclination
angle $\inclSK$, while the phase angle $\phaseSK$ is oscillating. After a few
tumbling motions, the capsule starts to perform stable swinging motions, where
the inclination angle $\inclSK$ oscillates around a stable mean value, while
the phase angle $\phaseSK$ is steadily decreasing. Independent studies using
two different solvers confirmed this finding: A transient motion relaxing from
a tumbling motion into a stable swinging motion was always observed in the
intermittent regime \cite{sui2008a, bagchi2009}. We therefore prefer to call
the disputed regime {transient} rather than intermittent
\cite{kessler2007}.

Close to the transition to tumbling, the oscillation of the inclination angle
$\inclSK$ is very strong, surprisingly leading to temporarily negative values
of the inclination angle, as first noted by Kessler \etal
\cite{kessler2007}. Here, the shape changes are very pronounced. This effect
was somewhat unfortunately named breathing by Bagchi \etal \cite{bagchi2009}.
It should not be confused with the vacillating-breathing motion for vesicles
\cite{misbah2006} because it does not show pronounced deformations
perpendicular to the shear plane. Moreover, the mean inclination angle is not
close to zero but finite. Since in all studies shape changes are highly
pronounced in the transient regime (see \cite{bagchi2009} in particular), this
reduced model has to be expanded to allow for shape changes to unambiguously
clarify the status of the disputed intermittency.

\section{Systematic quasi-spherical expansion} 
\label{sec:syst-quasi-spher}

From the numerical findings it is clear that allowing shape
changes is one important ingredient needed to solve the open
intermittency question. Noguchi \cite{noguchi2009a} incorporated
shape changes within the reduced model heuristically, following
extensions of the Keller-Skalak model for vesicles
\cite{noguchi2007}. Besides inclination and phase angle, a shape
parameter is used as a third degree of freedom. The elastic
energy due to shape changes is numerically calculated by
stretching a biconcave membrane (with a Skalak energy law). The
equation of motion of the shape parameter is deduced by a
deformation from the sphere, while the equations of motions for
the two angles are obtained by a Keller-Skalak derivation
\cite{keller1982}. The overall phase diagram of the extended
model was qualitatively no different from that of the reduced
model. In particular, the intermittent motions survived. A
temporarily negative inclination angle $\inclSK$ is also found in
the expanded model. The effect of the bending energy was studied
using the same heuristic approach \cite{noguchi2010b}.


\subsection{Equations of motion}
\label{sec:equations-motion-1}
We would like to replace this ad-hoc treatment of shape changes by a
systematic expansion of the full equations of motion.  Specifically, we expand
the equations of motion of the capsule for nearly spherical shapes. While the
deformation of the capsule shape from a sphere stays small, the displacement
of a membrane element during a tank-treading motion is usually large and
cannot be treated as a perturbation. The derivation of the elastic energy and
the equations of motion is straightforward, but would distract from the main
physics. Details are therefore presented in appendix~B
and
appendix~C.  From the elastic energy, we derive the elastic forces, which must
be balanced by hydrodynamic forces. The hydrodynamic forces have contributions
from the externally imposed flow and from the induced velocity field, which
must be solved dynamically. No-slip boundary conditions at the membrane and at
the external walls uniquely determine the dynamics.

Rather than specializing to linear shear flow, we consider general linear
external flow
\begin{equation}
  \Vv^{\infty} \equiv 
  s (x \vect{e}_{y} + y \vect{e}_{x})
  + \omega (x \vect{e}_{y} - y \vect{e}_{x}).
\end{equation}
Here, $s$ is a measure of the strength of the elongational flow
component and $2\omega$ is the vorticity. Linear shear flow of
the form $\Vv^{\infty}= \dot{\gamma} y \vect{e}_{x}$ corresponds to
$s=\dot{\gamma}/2$ and $\omega=-\dot{\gamma}/2$. In a
microfluidic four-mill device, which was used to study vesicle
dynamics quantitatively \cite{deschamps2009b}, it is possible to
control $\omega$ and $s$ independently.

For motions with the same mirror
symmetry with respect to the $xy$-plane as the external flow we
derive the equations of motion for the inclination angle
$\inclSK$, phase angle $\rotangle$, and a shape parameter
$\shape$. The shape parameter $\shape \in [0,\pi/2]$ is a measure
how much of the excess area is stored in the $m=2$ modes. For
$\shape=0$ the capsule is a prolate rotationally symmetric
ellipsoid with the long axis in the vorticity ($z$) direction.
For $\beta=\pi/3$ the capsule is again prolate rotationally
symmetric, but with the long axis in the shear ($xy$) plane.
$\beta$ is connected to the Taylor deformation parameter defined
in 
(\ref{eq:deform}) to lowest order via
\begin{equation}
  \deformSK \approx \frac {1}{4} \sqrt{\frac{15}{2\pi}\Delta}  \sin{\shape}.
\end{equation}
The corresponding shape parameter of the equilibrium shape is denoted with a
hat $\hat{\shape}$ (without loss of generality
$\hat{\inclSK}=\hat{\rotanglelab}=0$). The equations of motion as derived in
the appendices read
\begin{eqnarray}
  \label{eq:qseoma}\dot{\inclSK} + \dot{\rotangle} &=& - \Lambda\\
  \label{eq:qseomb}\dot{\rotangle} \sin\shape  &=& -S^{-1}\sin 2\rotangle -
  \cos 2\inclSK\\
  \label{eq:qseomc}\dot{\shape} &=& - S^{-1} \cot \hat{\shape} \sin\shape +
  \cos \shape (S^{-1} \cos2\rotangle+\sin2\inclSK),
\end{eqnarray}
with
\begin{eqnarray}
  \label{eq:paramsevans}
  \tau &\equiv&
  8 \sqrt{\frac{30\pi}{\Delta}}
  \frac{s}{32+23\etain/\etaout} t,\\
  S &\equiv&
  \sqrt{\frac{30\pi}{\Delta}}
  \frac{R\etaout}{\mu \sin\hat\shape} s,\\
  \Lambda &\equiv& -
  \frac 1 8 \sqrt{\frac{\Delta}{30\pi}}
  (32+23\etain/\etaout) \frac{\omega}{s}.
\end{eqnarray}
Here $\tau$ denotes a dimensionless time and the dot the
corresponding derivative. $S$ is a measure of the
strength of the elongational flow, and $\Lambda$ measures the
relative strength of the rotational flow. In linear shear flow
($s=-\omega=\dot{\gamma}/2$), $\Lambda$ depends only on the
viscosity contrast and material capsule parameters, while $S$ is
proportional to the shear rate $\dot{\gamma}$.

\subsection{Numerical results}
\label{sec:numerical-results}
\begin{figure}
  \centering
  \includegraphics[width=0.49\textwidth]{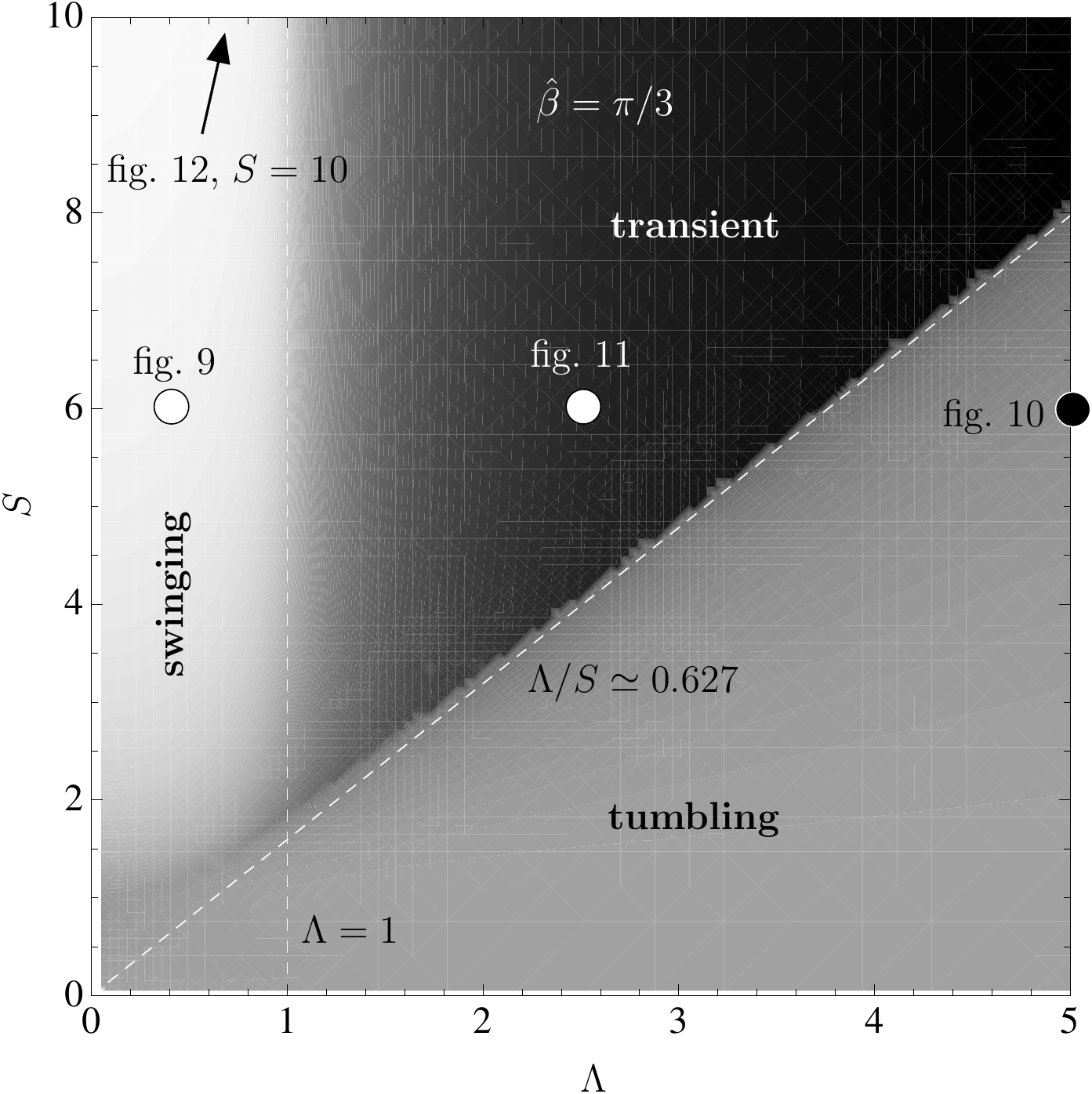}
  \caption{Dynamical phase diagram obtained by numerically integrating
    (\ref{eq:qseoma}--\ref{eq:qseomc}) for an initially axisymmetric
    equilibrium capsule ($\shape(\tau=0)=\hat\shape=\pi/3$). The mean shape
    parameter $\langle \shape \rangle$ is shown in a grey scale plot as a
    function of the shear rate $S$ and the viscosity contrast $\Lambda$. Dark
    grey corresponds to the minimal value $\langle \shape \rangle=0$, while
    light grey denotes a maximal value $\langle \shape \rangle=\pi/2$. Three
    regimes can be distinguished: swinging, tumbling and a transient motion
    towards swinging. The small circles correspond to
    figures~\ref{fig:inclinationa}--\ref{fig:inclinationc}, while the cut at
    $S=10$ through the phase diagram is shown in
    figure~\ref{fig:cutshapea}. The vertical dashed line corresponds to the
    analytically obtained phase boundary at $\Lambda=1$ between swinging and
    transient motion in the limit $S\to\infty$ with constant $\Lambda$. The
    inclined dashed white line corresponds to the phase boundary between
    tumbling and transient motion in the limit $S,\Lambda\to\infty$ with
    constant ratio $\Lambda/S\simeq 0.627$.}
  \label{fig:tumblingrate}
\end{figure}

\begin{figure}
  \centering
  \includegraphics[width=0.46\textwidth]{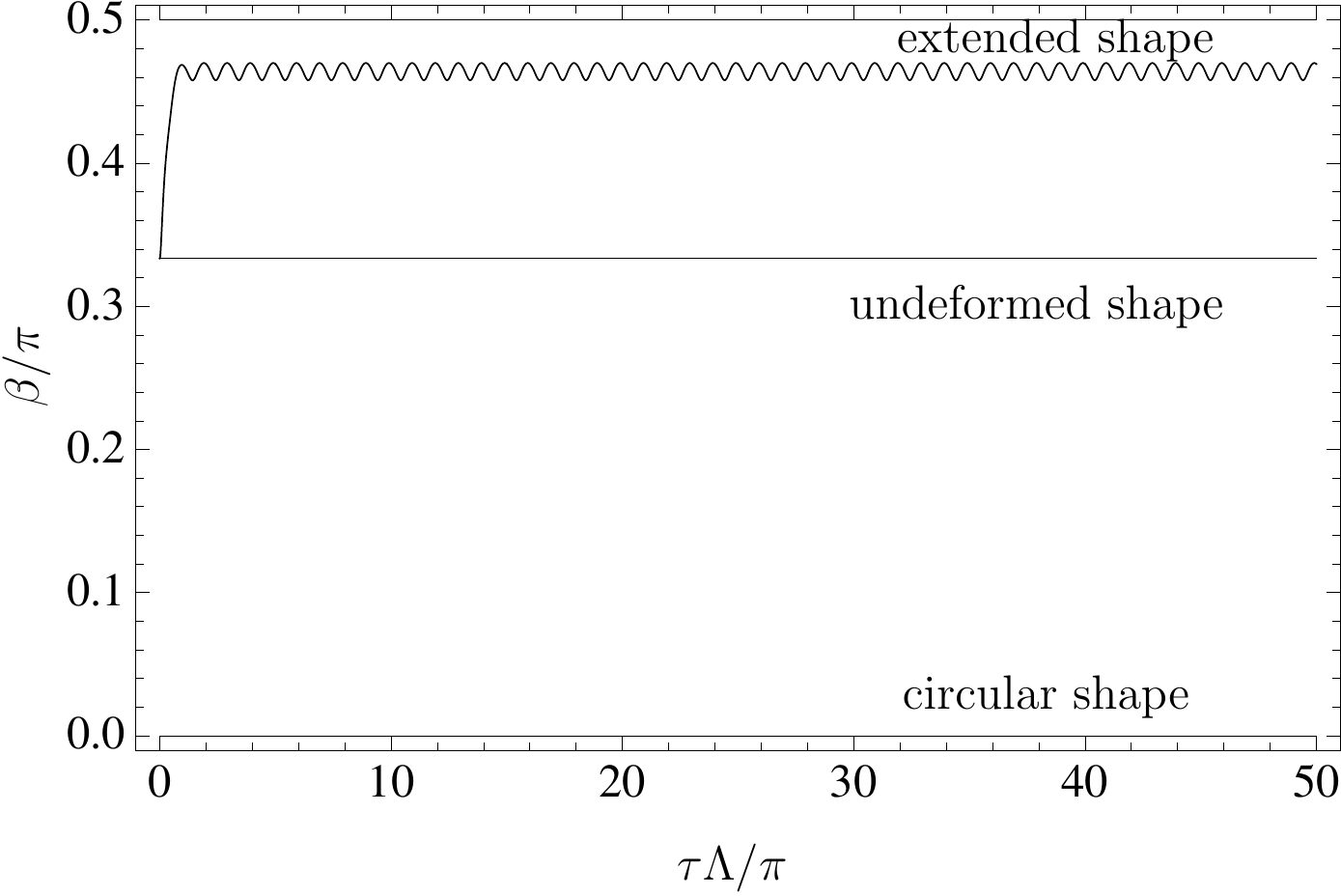}
  \includegraphics[width=0.52\textwidth]{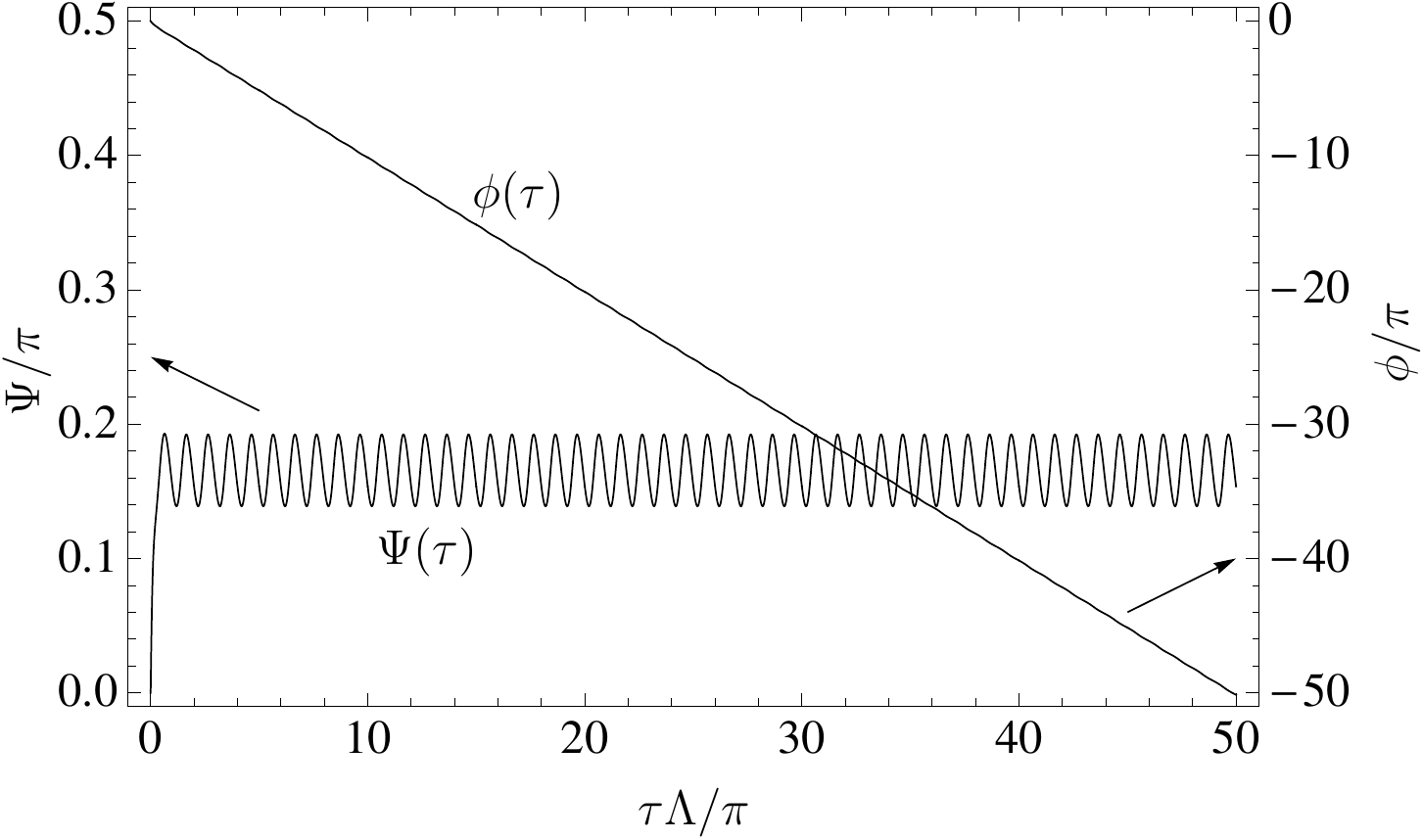}
  \caption{Shape parameter $\shape$ (left), inclination angle
    $\inclSK$, and phase angle $\phaseSK$ (right) as a function
    of dimensionless time $\tau\Lambda/\pi$ within the swinging
    regime ($\hat\shape=\pi/3$, $\Lambda=0.5$ and $S=6.0$, see
    figure~\ref{fig:tumblingrate}). The shape parameter $\shape$
    oscillates around a value $\langle\shape\rangle$ close to
    $\pi/2$, implying an almost maximally extended shape within
    the shear plane. The inclination angle $\inclSK$ oscillates
    around a constant positive value $\langle\inclSK\rangle$,
    while the phase angle $\phaseSK$ linearly decreases with
    superimposed oscillations (not seen in the plot due to scale). }
  \label{fig:inclinationa}
\end{figure}

\begin{figure}
  \centering
  \includegraphics[width=0.46\textwidth]{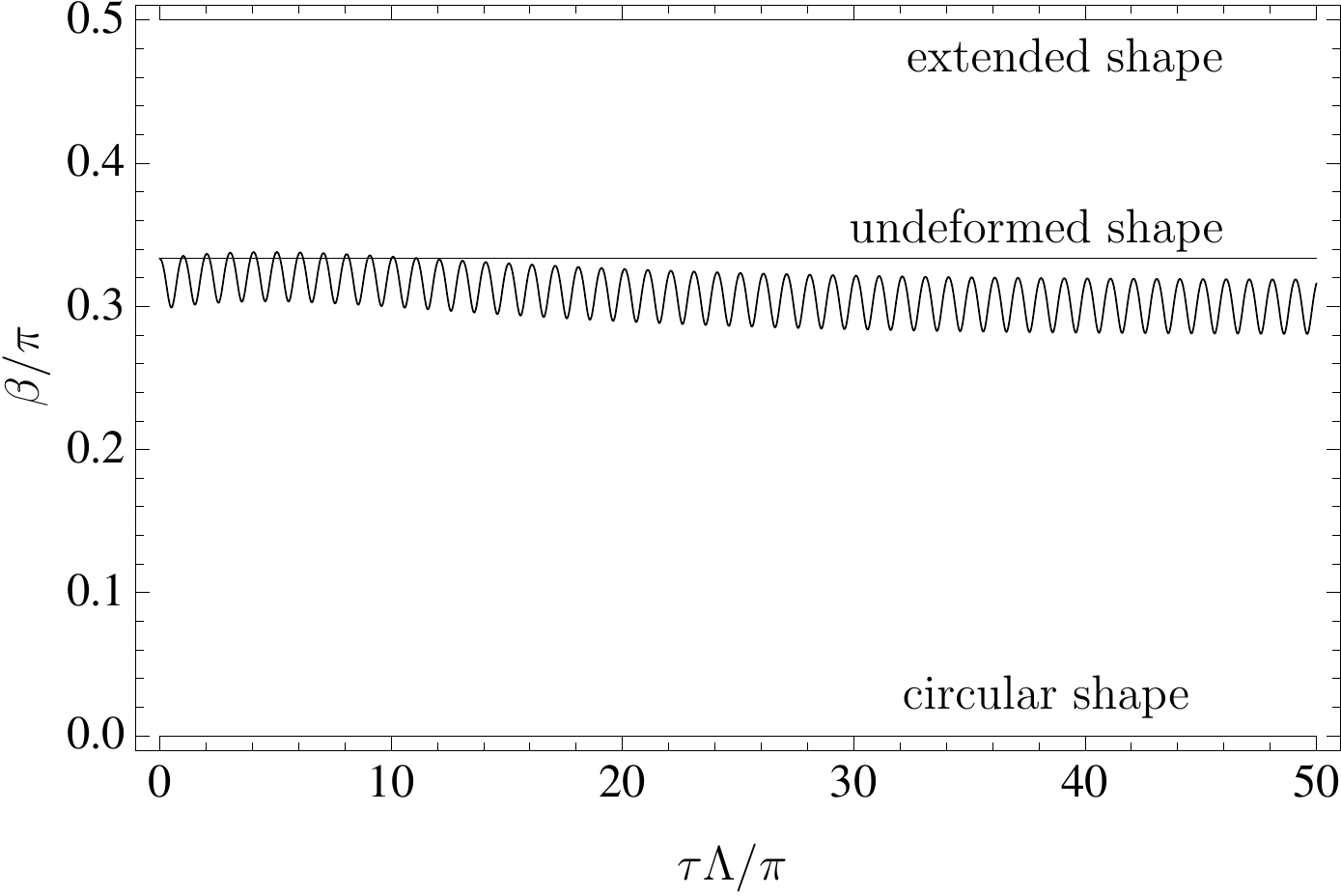}
  \includegraphics[width=0.52\textwidth]{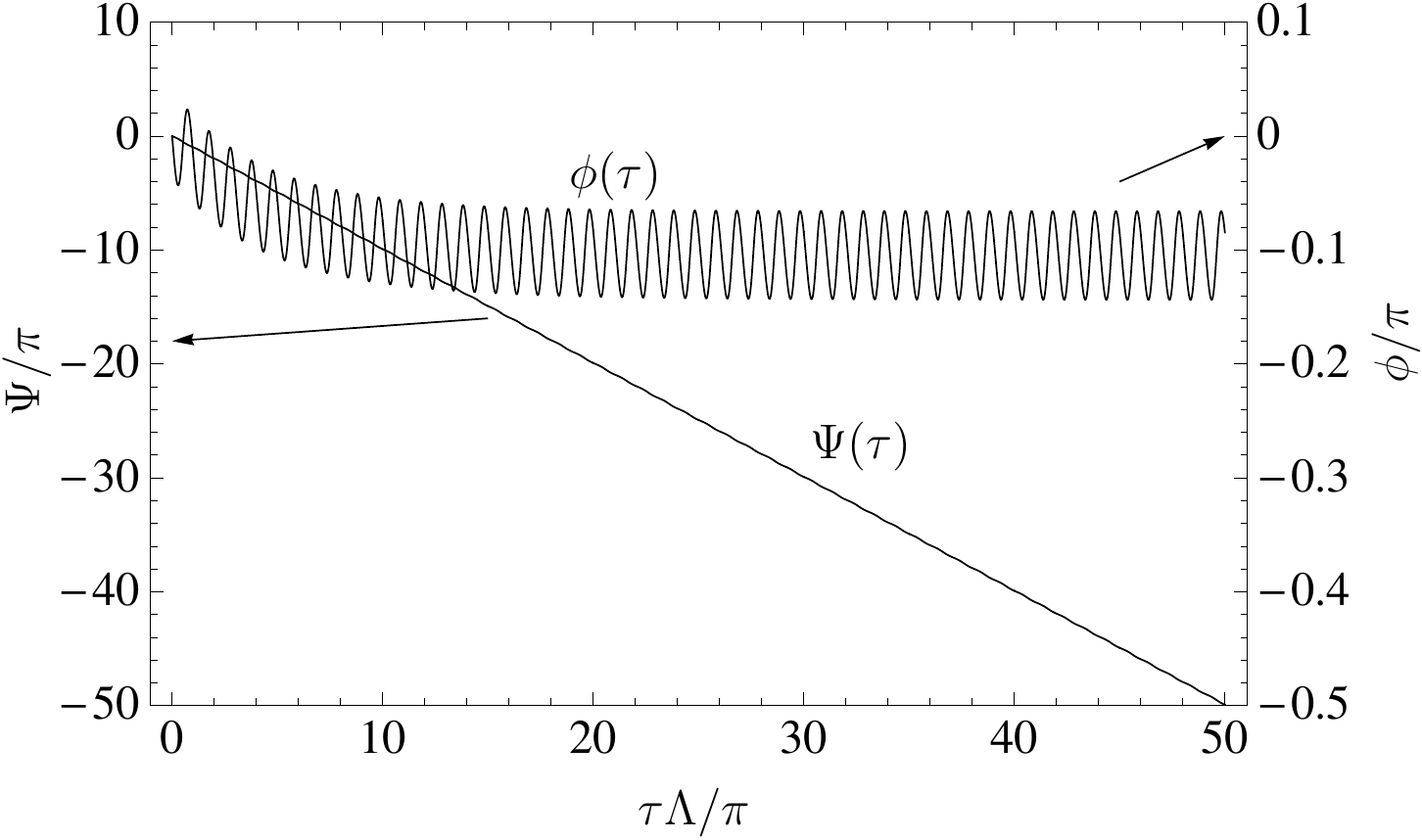}
  \caption{Shape parameter $\shape$ (left), inclination angle
    $\inclSK$, and phase angle $\phaseSK$ and as a function of
    dimensionless time $\tau\Lambda/\pi$ within the tumbling
    regime ($\hat\shape=\pi/3$, $\Lambda=5.0$ and $S=6.0$, see
    figure~\ref{fig:tumblingrate}). The shape parameter $\shape$
    oscillates around a value close to its reference value
    $\hat\shape=\pi/3$. The inclination angle $\inclSK$ linearly
    decreases with superimposed oscillations, while the phase
    angle $\phaseSK$ oscillates around a constant positive value
    $\langle\phaseSK\rangle$. }
  \label{fig:inclinationb}
\end{figure}

\begin{figure}
  \centering
  \includegraphics[width=0.46\textwidth]{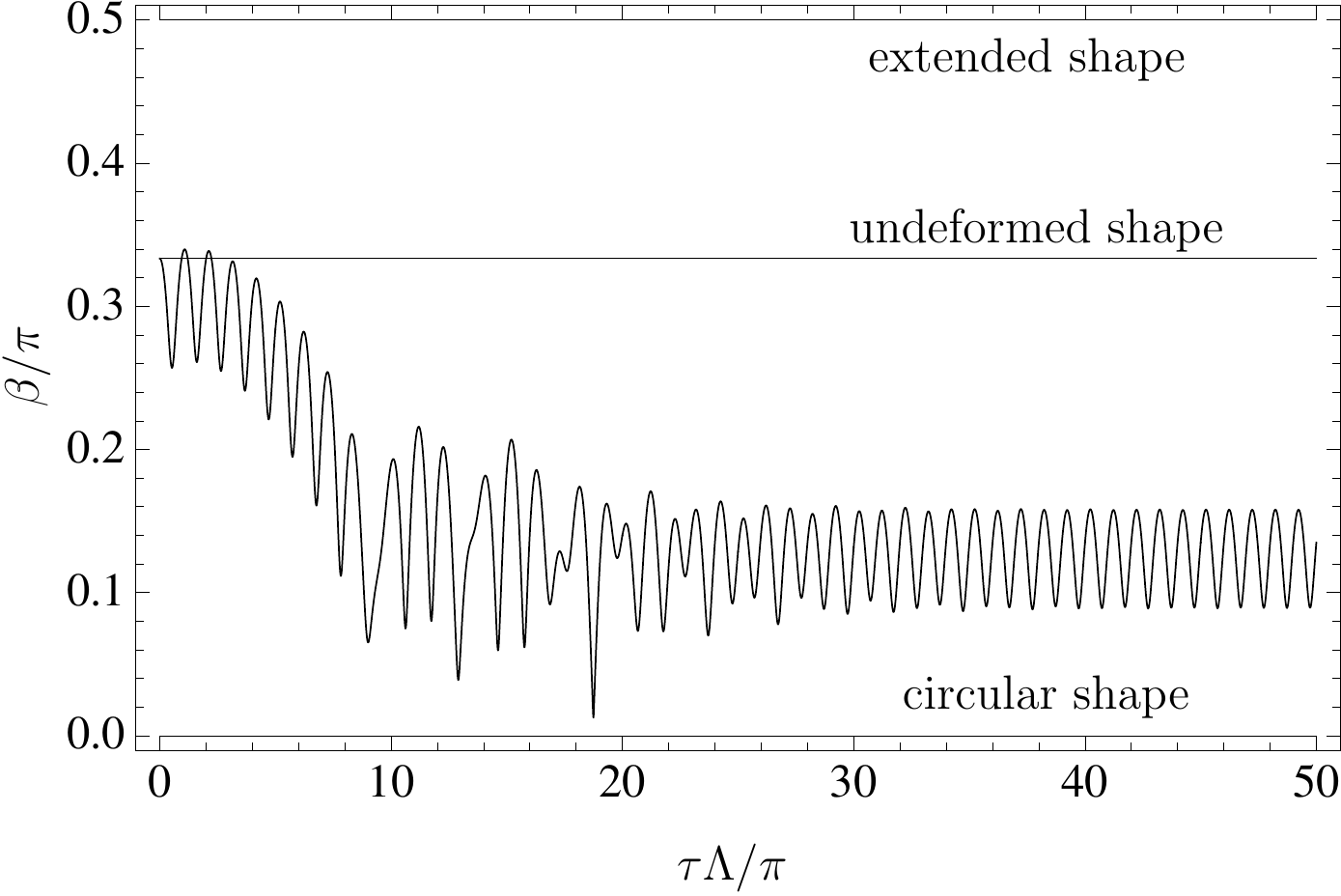}
  \includegraphics[width=0.52\textwidth]{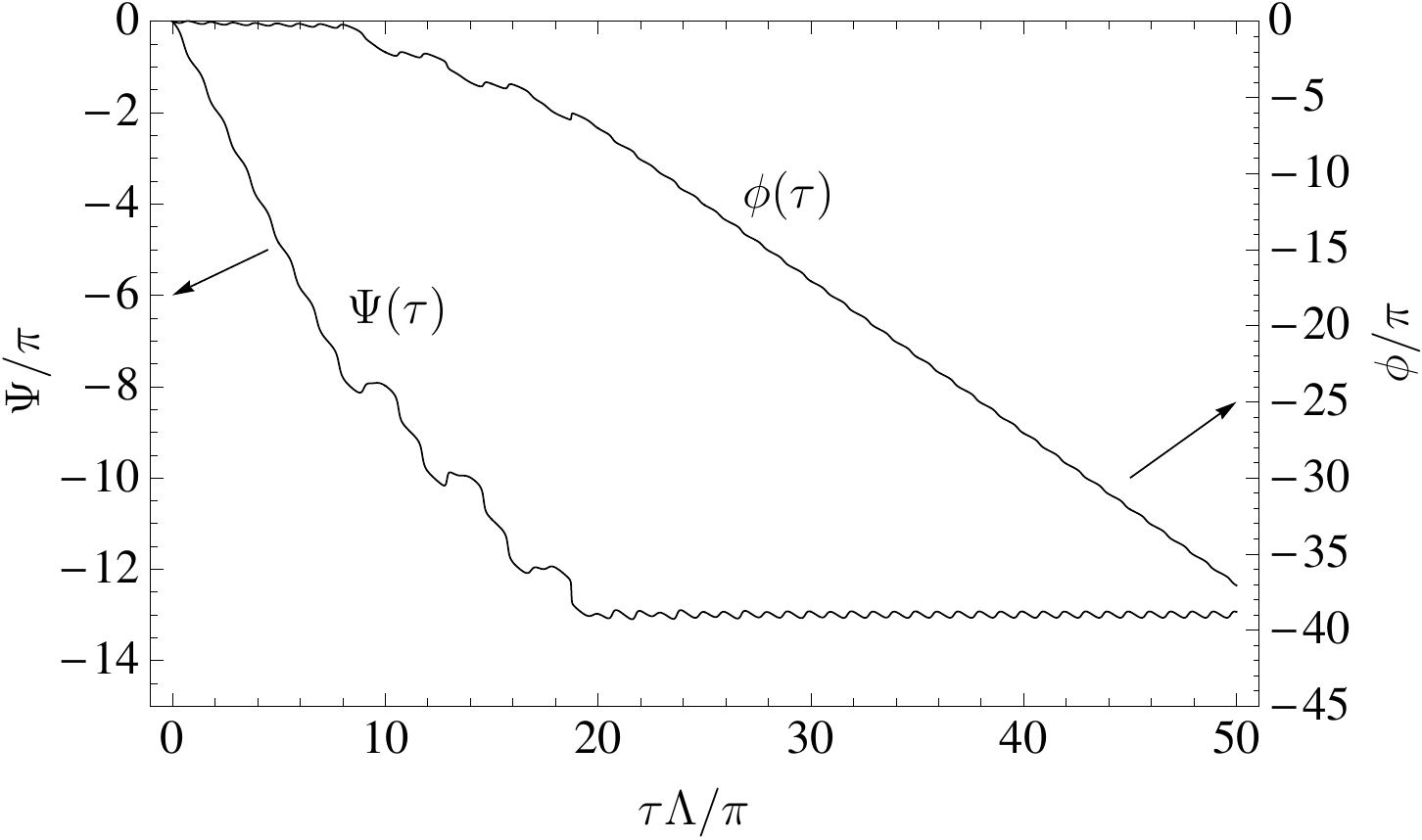}
  \caption{Shape parameter $\shape$ (left), inclination angle
    $\inclSK$, and phase angle $\phaseSK$ (right)
     as a function of dimensionless time $\tau\Lambda/\pi$
    within the transient regime ($\hat\shape=\pi/3$, $\Lambda=2.5$ and
    $S=6.0$, see figure~\ref{fig:tumblingrate}). Initially, tumbling
    motions occur where the inclination angle $\inclSK$ linearly decreases
    with superimposed oscillations, while the phase angle $\phaseSK$
    oscillates around a constant positive value. On a longer time scale, the
    shape parameter $\shape$ relaxes to a smaller value, where the
    capsule is close to circular in the shear plane and strongly extended in
    the vorticity direction. After the relaxation, the motion changes into a stable swinging,
    where oscillations are superimposed to a constant mean inclination angle
    $\langle\inclSK\rangle$ and a linear decreasing phase angle $\phaseSK$.}
  \label{fig:inclinationc}
\end{figure}

Before we proceed to analyze these equations in more detail, we summarize the
findings based on numerically integrating these equations for an initially
axisymmetric equilibrium capsule with $\shape(0)=\hat{\shape}=\pi/3$. In
figure~\ref{fig:tumblingrate}, the mean shape parameter $\langle\shape\rangle$
in the long time limit is plotted colour coded as a function of the viscosity
contrast $\Lambda$ and the elongational shear rate $S$.  In this ``dynamical
phase diagram'', one can distinguish three regions: (i) A swinging region,
roughly bounded by $\Lambda < 1$ and $S>1$.  The mean tumbling rate
$\tumbnormalized\simeq0$ vanishes in this region, while the shape parameter
$\beta\simeq\pi/2$ is maximal.  A typical trajectory over time is shown in
figure~\ref{fig:inclinationa}. (ii) A tumbling region roughly bounded by
$\Lambda > 0.627 S$. Here, $\tumbnormalized \simeq 1$ takes the highest
possible value, while the shape parameter oscillates around the equilibrium
one, $\shape\simeq\hat{\shape}$, in the developed tumbling regime. A typical
trajectory is shown in figure~\ref{fig:inclinationb}. (iii) A transient region
in between. Here, the shape parameter $\shape \simeq 0$ is small, while the
tumbling rate $\tumbnormalized\simeq 0$ vanishes in the long time
limit. A typical trajectory shown in figure~\ref{fig:inclinationc} reveals that
the capsule initially tumbles for a while, before it assumes a stationary
swinging state. The initial tumbling time increases with increasing $\Lambda$.

\begin{figure}
  \centering
  \includegraphics[width=0.74\textwidth]{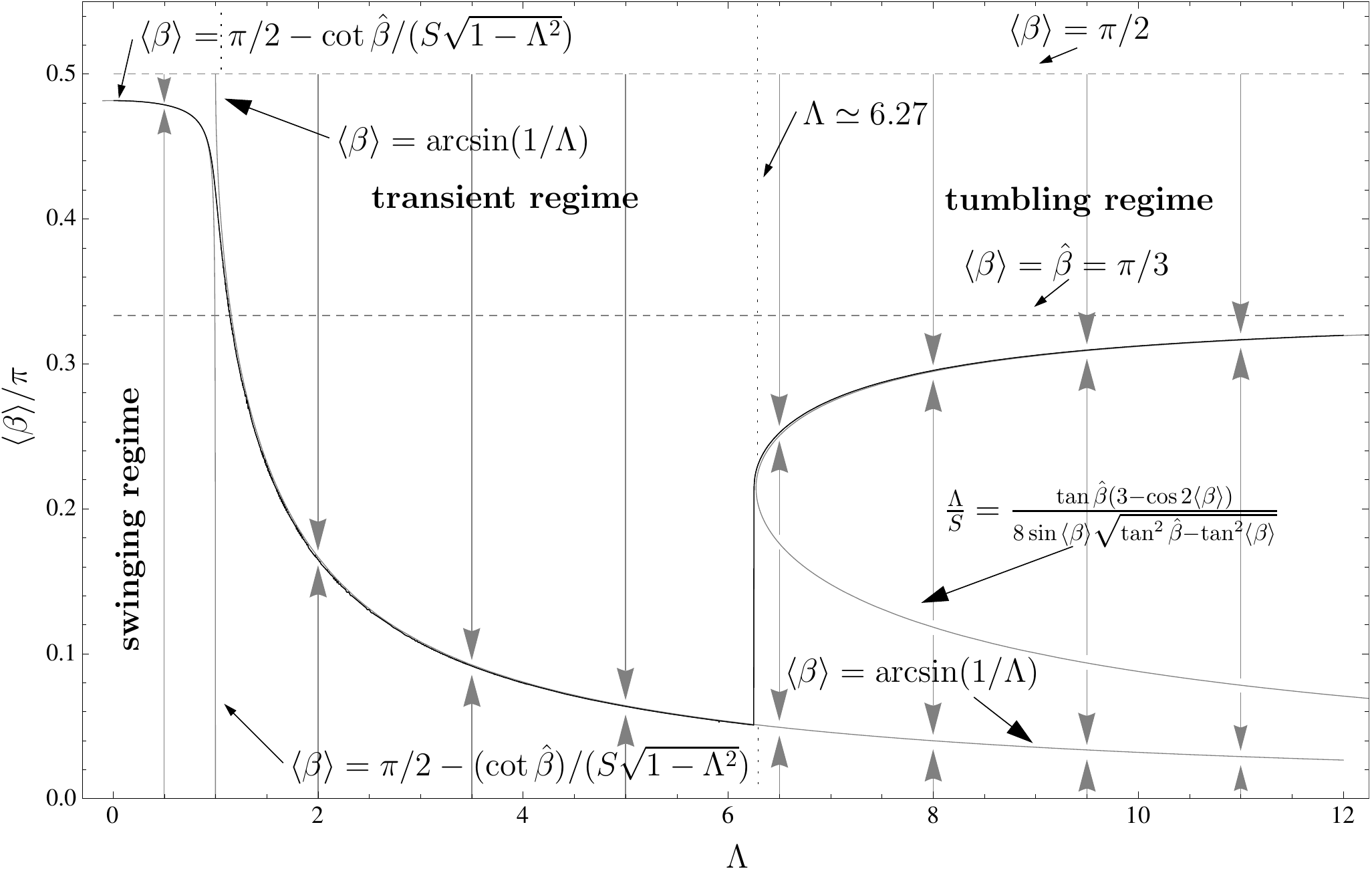}
  \caption{Cut through the phase diagram of figure~\ref{fig:tumblingrate}. The
    numerically obtained mean shape parameter $\langle\shape\rangle$ is plotted as a
    function of the viscosity contrast $\Lambda$ at constant shear rate $S=10$
    for an axisymmetric capsule starting out of its reference shape
    $\shape(0)=\hat{\shape}=\pi/3$ (black line). The grey lines correspond
    to asymptotic analytical results. For low $\Lambda<1$ the capsule swings
    with maximally extended shape within the shear plane. For high values
    $\Lambda>6.27$ the capsule performs tumbling motions with shapes close to
    the reference shape. In between there is the transient regime where the
    extension of the shape in the vorticity direction strongly increases with
    increasing $\Lambda$. The arrows denote the flow towards the long-time
    stationary value.}
  \label{fig:cutshapea}
\end{figure}

The shape parameter $\shape$ can therefore act as an order parameter for
distinguishing the dynamical phases. A cut through the dynamical phase diagram
at $S=10$, initially starting at the equilibrium shape
$\shape(t=0)=\hat{\shape}$ is shown in figure~\ref{fig:cutshapea}. We see that
the swinging to tumbling transition at $\Lambda\simeq 1$ is continuous, while
$\shape$ jumps at the transient to tumbling transition at $\Lambda\simeq
6.27$, like in a first order phase transition.

\subsection{Analytical results}\label{sec:analytical-results}
After having presented the numerical phenomenology, we analyze the equations
of motion analytically and make the connection to the reduced model of
section~\ref{sec:reducedmodel}, where the shape $\shape$ of the capsule is
held fixed. If we ignore the dynamics (\ref{eq:qseomc}) of the shape parameter
and compare the remaining equations~(\ref{eq:qseoma}, \ref{eq:qseomb}) with
the corresponding ones in the reduced model, (\ref{eq:dim1}, \ref{eq:dim2}),
we see that up to a rescaling in time both equations are identical in the
quasi-spherical limit ($\alpha\to 0$). We identify
\begin{eqnarray}
  \tau_{\text{SK}} &\equiv& \tau_{\text{QS}} / \sin(\shape),\\
  \lambda &\equiv& \Lambda \sin(\shape),\\
  \chi&\equiv& S.
\end{eqnarray}
Thus, if the dynamics of the shape parameter $\shape$ is
slow compared to that of the angles, we can read off the fast
capsule dynamics in the phase diagram of the reduced model, where
the effective viscosity ratio $\lambda$ changes slowly in time.
We will see that for strong flows the shape changes indeed on a
much slower time scale than the angles. This separation of time
scales is the key to understanding the dynamic phase behaviour of
capsules in the quasi-spherical description.

\subsubsection{Swinging/transient transition, $S \gg 1$, $\Lambda$ fixed:}
We start our analysis by taking the limit of strong flow $S \to
\infty$, while keeping the relative strength of the rotational
flow $\Lambda$ constant. The leading order equations then become
\begin{eqnarray}
\label{eq:swinginga}\dot{\inclSK} + \dot{\rotangle} &=& - \Lambda,\\{}
 (\dot{\inclSK} + \Lambda) \sin \shape &=&  \cos 2\inclSK,\\
\label{eq:swingingc}\dot{\shape} &=&   \cos \shape \sin 2\inclSK.
\end{eqnarray}
We can identify two dynamical regimes.

\begin{figure}[ht]
  \centering
  \includegraphics[width=0.74\linewidth]{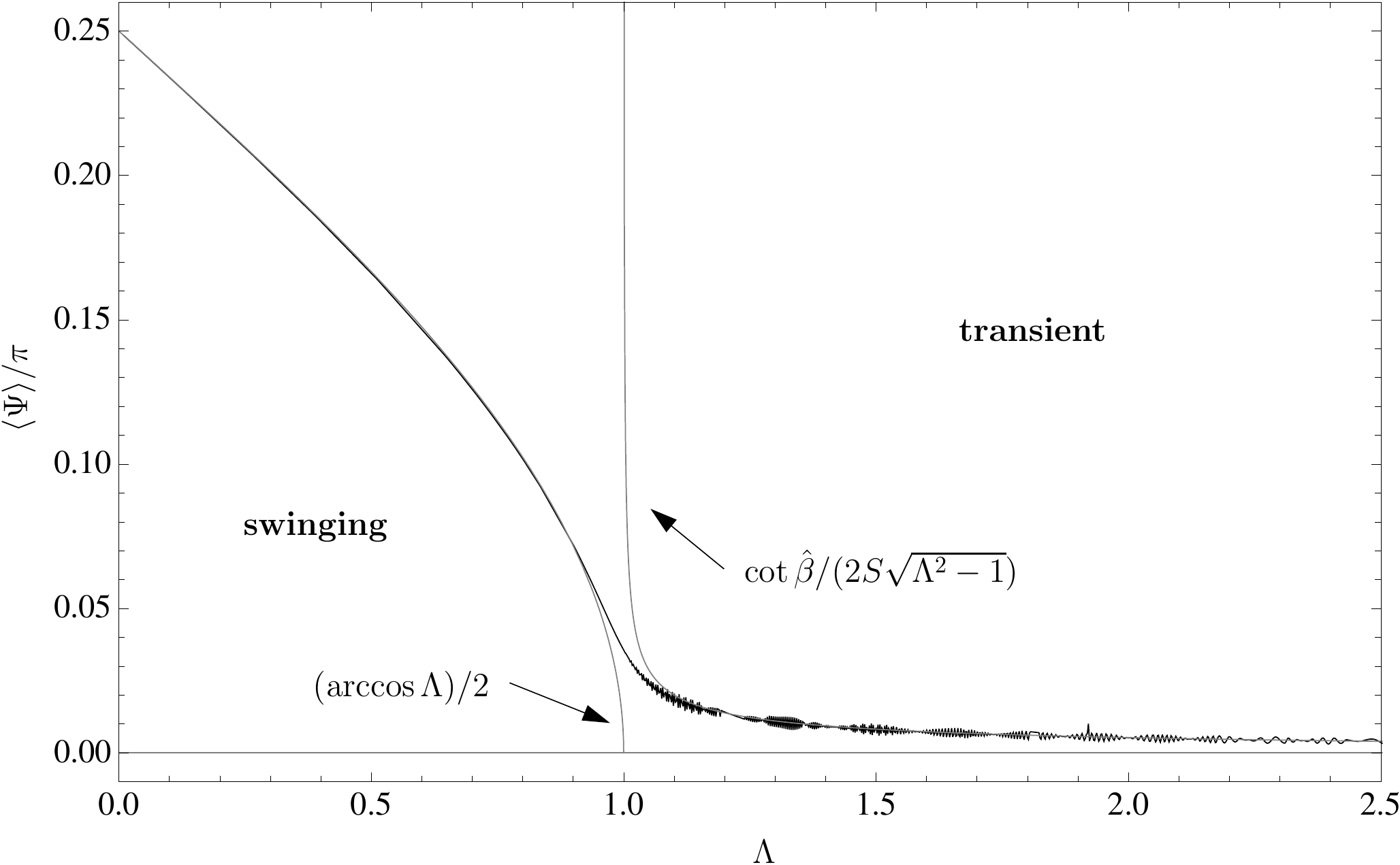}
  \caption{Mean inclination angle $\langle \inclSK \rangle$ as a function of
    the viscosity ratio $\Lambda$ for $\hat\shape=\pi/3$ and $S=10$. The black
    line corresponds to the numerically obtained values, while the grey lines
    show analytical results. The mean inclination angle decreases from $\pi/4$
    (corresponding to the direction of the elongational part of the shear flow) at
    $\Lambda=0$ in the swinging regime. At the transition to the transient
    regime, the mean inclination angle is close to $0$. In the
    transient regime, $\langle \inclSK \rangle$ asymptotically reaches $0$.}
  \label{fig:inclmean}
\end{figure}

\begin{figure}[ht]
  \centering
  \includegraphics[width=0.74\linewidth]{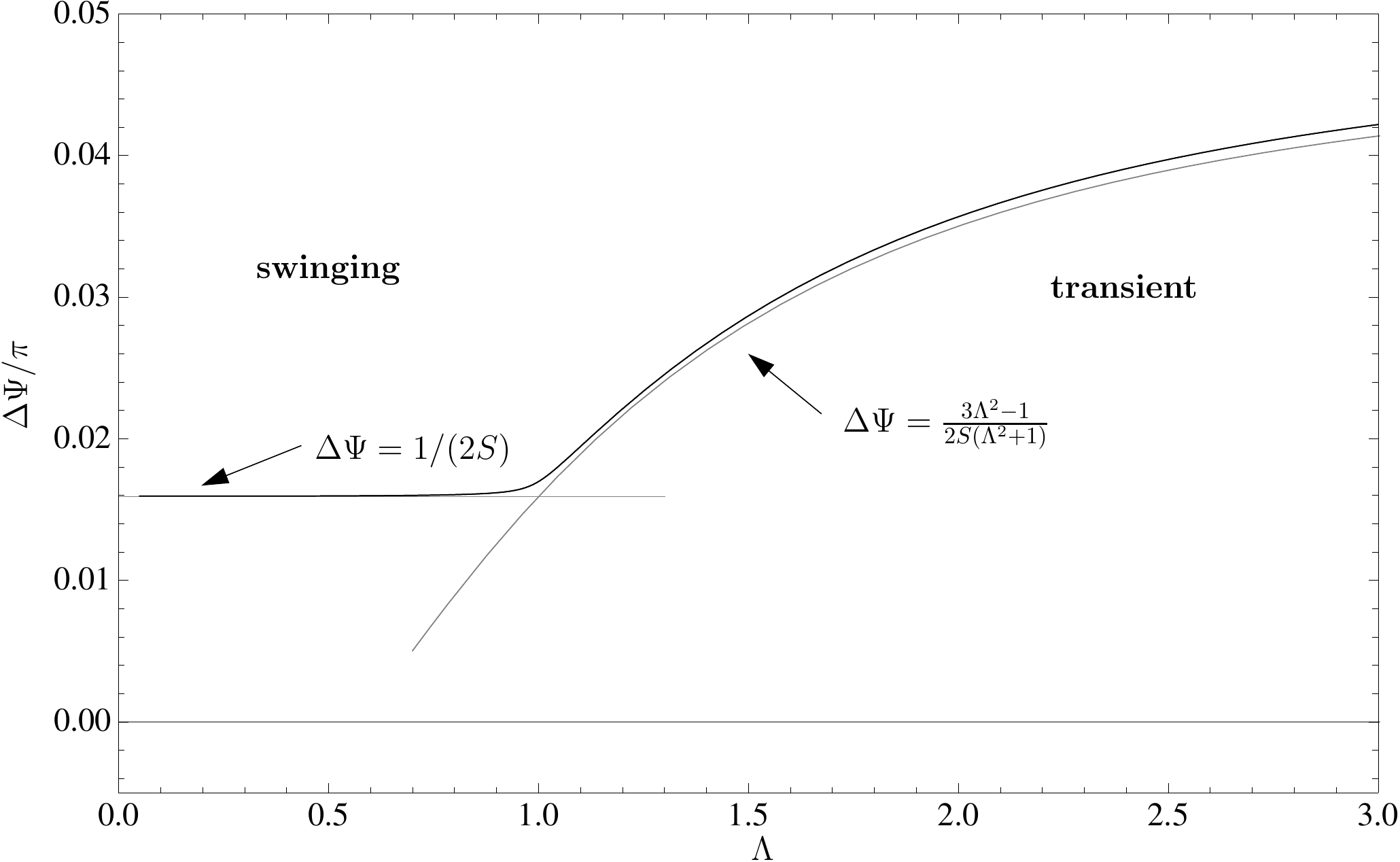}
  \caption{Amplitude $\Delta\inclSK$ of the oscillations of the inclination
    angle $\inclSK$ as a function of the viscosity ratio $\Lambda$ for
    $\hat\shape=\pi/3$ and $S=10$. The black line corresponds to the
    numerically obtained values, while the grey lines show analytical
    results. }
  \label{fig:incloszi}
\end{figure}

\begin{figure}[ht]
  \centering
  \includegraphics[width=0.74\linewidth]{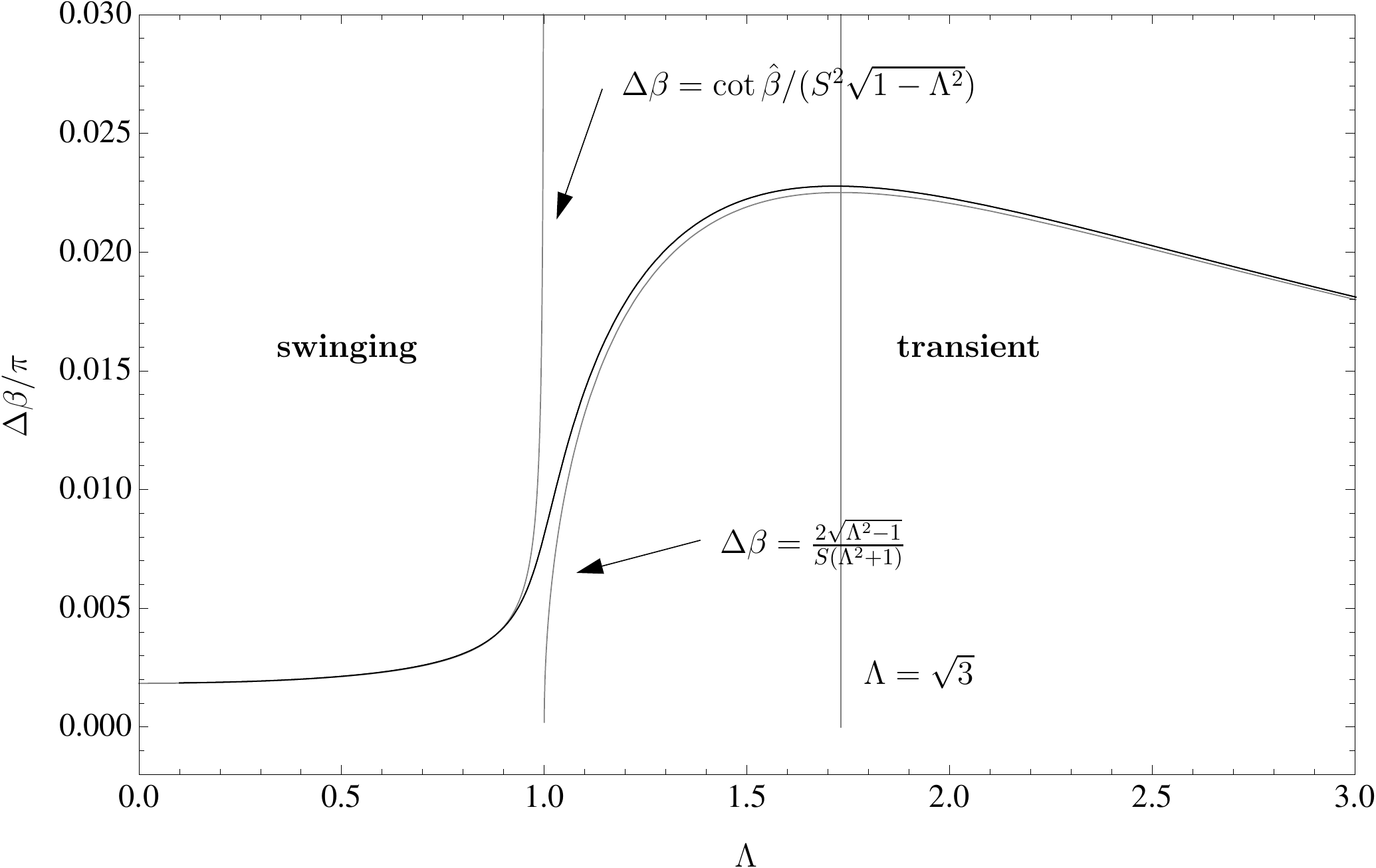}
  \caption{Amplitude $\Delta\shape$ of the oscillations of the shape parameter
    $\shape$ as a function of the viscosity ratio $\Lambda$ for
    $\hat\shape=\pi/3$ and $S=10$. The black line corresponds to the
    numerically obtained values, while the grey lines show analytical
    results. In the swinging regime, the oscillations show up only in second
    order in $1/S$, while in the transient regime, the amplitudes are of order
    $1/S$. Shape changes of the swinging capulse are small in the swinging
    regime and pronounced in the transient regime.}
  \label{fig:shapeoszi}
\end{figure}

\emph{Swinging: ($\Lambda < 1$):}
We now look for stationary tank-treading solutions for the
inclination angle $\inclSK = \inclSK_{0}$ and the shape parameter
$\shape=\shape_{0}$.
Assuming that $\sin 2\inclSK$ stays positive ($0 < \inclSK_{0} <\pi/2$), the shape
parameter approaches the value $\shape = \pi/2$ asymptotically.
The remaining two equations then reduce to the Keller-Skalak form
  $\dot{\inclSK} = - \Lambda + \cos 2\inclSK $
with the stationary solution for $\Lambda < 1$
\begin{equation}
  \inclSK_{0}^{\swSCR} = \frac{1}{2} \arccos\Lambda.
\end{equation}
The next order in the inverse shear rate yields oscillating
perturbations of the inclination angle 
\begin{equation}
  \label{eq:inclsw}
  \inclSK^{\swSCR} = \frac{1}{2}  \arccos \Lambda
  + \frac 1 {2S} \cos 2 \Lambda \tau + \mathcal{O}(1/S^{2}),
\end{equation}
while the shape parameter is shifted with second order oscillations
\begin{equation}
\label{eq:shapesw}
\shape^{\swSCR} = \frac \pi 2 - \frac{\cot\hat\shape}{S
  \sqrt{1-\Lambda^2}} + \frac{\cot\hat\shape}{S^{2}
  \sqrt{1-\Lambda^2}} \sin 2\Lambda \tau + \mathcal{O}(1/S^{3}).
\end{equation}
These results correspond to the numerically observed swinging dynamics. In
figure~\ref{fig:inclmean}, the numerically calculated long term mean
inclination angle $\langle \inclSK \rangle$ is compared to the corresponding
first term on the right hand side of
(\ref{eq:inclsw}) as a function of $\Lambda$. The amplitude of the oscillatory
component, the second term on the right hand side of
(\ref{eq:inclsw}), is compared to numerically obtained data in
figure~\ref{fig:incloszi}. The amplitude of the second order oscillations of
the shape parameter $\shape$ finally is shown in figure~\ref{fig:shapeoszi}.

\emph{Transient ($\Lambda > 1$):} For $\Lambda>1$
(\ref{eq:swinginga}--\ref{eq:swingingc}) admit a different stationary solution
given by
\begin{equation}
  \inclSK_{0}^{\trSCR} =0 \qquad \text{and} \qquad
  \shape_{0}^{\trSCR} = \arcsin(1/\Lambda).
\end{equation}
These values define the stationary state in the transient regime, which
is only reached after initial tumbling motions.  The first order calculations add 
oscillatory terms according to
\begin{eqnarray}
\rotangle^{\trSCR} &=& -\Lambda \tau - \frac{\cot\hat\shape}{2S\sqrt{\Lambda^2-1}}
- \frac{3\Lambda^2-1}{2S(\Lambda^2+1)} \cos 2\Lambda \tau+ \mathcal{O}(1/S^{2}),\\
\inclSK^{\trSCR} &=& \frac{\cot\hat\shape}{2S\sqrt{\Lambda^2-1}}
+ \frac{3\Lambda^2 - 1}{2S(\Lambda^2+1)} \cos 2\Lambda \tau + \mathcal{O}(1/S^{2}),\\
\shape^{\trSCR} &=& \arcsin 1/\Lambda
+ \frac{2\sqrt{\Lambda^2-1}}{S(\Lambda^2+1)} \sin 2\Lambda \tau + \mathcal{O}(1/S^{2}).
\end{eqnarray}
Again, the mean value of the inclination angle and the
oscillation amplitudes of the inclination angle and shape
parameter are compared to numerical data in figures
\ref{fig:inclmean}--\ref{fig:shapeoszi}.

\subsubsection{Transient/tumbling transition, $S, \Lambda \gg 1, \Lambda/S$
  fixed:}
In order to discuss the tumbling regime, we must consider a different
asymptotic limit. The numerical results indicate that the transient to
tumbling transition occurs for large shear rates at fixed ratio
$\Lambda/S$. In this limit, two timescales must be considered. On average the
parameters vary on a slow timescale $\tau$, but the fast oscillating
trigonometric terms produce oscillations on the much faster timescale $\Lambda
\tau$. We seek an expansion of $\inclSK$ ($\rotangle$, $\shape$) into powers
of the small quantity $1/\Lambda$, $\inclSK \equiv \inclSK_{0} + \inclSK_{1} +
\ldots$. The time derivative of the small oscillating corrections
$\inclSK_{1}$ are multiplied by $\Lambda$ and can thus contribute to the order
one on the left hand side of the equations of motion. The systematic method to
analyze processes with different time scales in a perturbation scheme is
described in \cite{hinch1991}.

\emph{Long time swinging behaviour:} First, we seek to rediscover the
stationary tank-treading solution in the transient regime. We expect
$\dot{\rotangle} \simeq -\Lambda$ to leading order. Since only terms of order
one occur on the right hand side of
(\ref{eq:qseomb}), $\sin \shape$ must be small of order $1/\Lambda$ on the
left hand side. Taking this into account, we recover the tank-treading
solution ($\arcsin(1/\Lambda)\approx 1/\Lambda$)
\begin{equation}
   \inclSK^{\swSCR}_{0} = 0\qquad \text{and} \quad  \beta^{\swSCR}_{0} = 1/\Lambda.
\end{equation}
Expanding the equations of motion
(\ref{eq:qseoma}--\ref{eq:qseoma}) to the next orders in $1/S$
gives the oscillatory corrections
\begin{eqnarray}
   \inclSK^{\swSCR} &=& \frac{3}{2S} \cos 2 \Lambda \tau + \mathcal{O}(1/S^{2}),\\
   \rotangle^{\swSCR} &=& - \Lambda \tau - \frac{3}{2S} \cos 2
   \Lambda \tau + \mathcal{O}(1/S^{2}),\\
  \beta^{\swSCR} &=& 1/\Lambda + \frac{2}{\Lambda S}\sin 2 \Lambda
  \tau + \mathcal{O}(1/S^{3}).
\end{eqnarray}
Note that the swinging motion is linearly stable for arbitrarily
high values of $\Lambda$.

\emph{Tumbling:}
During tumbling motion the inclination angle $\inclSK$ grows very fast, to
leading order $\dot{\inclSK} \simeq -\Lambda$. We assume a slow
dynamics of the shape $\shape_{0}(\tau)$, superimposed by small
but fast oscillations $\shape_{1}(\Lambda \tau)$. From the
leading order equations of 
(\ref{eq:qseoma}--\ref{eq:qseomc}) we obtain
\begin{eqnarray}
  \rotangle_{1}(\tau) &=& - \frac{1}{2\Lambda\sin \shape_{0}}
  \sin(2\Lambda \tau),\\
  \shape_{1}(\tau) &=&  \frac{\cos \shape_{0}}{2\Lambda} \cos 2\Lambda \tau.
\end{eqnarray}
The equations at order $1/\Lambda$ give
\begin{eqnarray}
  \label{eq:tumblingeoma}
  \dot \rotangle_{0} &=& - \frac{1}{S}
  \frac{\sin 2\rotangle_{0}}{\sin \shape_{0}} - \left\langle
    \frac{\cos 2\Lambda \tau}{\sin\shape_{0}+\shape_{1}}
  \right\rangle 
  \\ \nonumber
  &=& - \frac{1}{S}
  \frac{\sin 2\rotangle_{0}}{\sin\shape_{0}} -
  \frac{1}{4\Lambda}
  \frac{1+\sin^{2}\shape_{0}}{\sin^{2}\shape_{0}}\\
  \label{eq:tumblingeomc}  \dot \shape_{0} &=& - \frac{1}{S} \cot \hat{\shape}
  \sin\shape_{0} + \frac{1}{S}
  \cos\shape_{0}\cos 2\rotangle_{0} + \frac{1}{S} \langle \cos
  \shape \sin2\inclSK \rangle,
\end{eqnarray}
where $\langle\cdots\rangle$ denotes averaging over one period of the fast
oscillations. The average of the cross term $\langle \cos \shape \sin
2\inclSK\rangle$ on the right hand side of
(\ref{eq:tumblingeomc}) is zero, since the oscillations of $\shape$ and
$\inclSK$ have a relative phase shift of $\pi$.

\emph{Stable tumbling regime:}
Equations~(\ref{eq:tumblingeoma}--\ref{eq:tumblingeomc}) admit a stationary
solution determined by the conditions
\begin{eqnarray}
\label{eq:tumblingstata}  \sin 2\rotangle_{0} &=& -
  \frac{1+\sin^{2}\shape_{0}}{\sin\shape_{0}} \frac{S}{4\Lambda},\\
\label{eq:tumblingstatb}  \cos 2\rotangle_{0} &=& \frac{\tan
    \shape_{0}}{\tan\hat{\shape}}.
\end{eqnarray}
The solution $\shape(\Lambda/S)$ of this system, which is shown
in figure~\ref{fig:cutshapea}, has two branches. The upper branch
was found to be numerically stable. For $\Lambda/S\to \infty$ the
stable branch approaches $\beta_{0}\to\hat\beta$ (with
$\rotangle_{0}\to 0$). On the unstable branch we have
$\beta_{0}\to 0$ (with $\rotangle_{0} \to \pi/4$) as
$\Lambda/S\to \infty$. We can avoid multi-valued functions by
instead writing $\Lambda/S$ as a function of $\shape_{0}$
\begin{equation}
  \label{eq:lambdacrit}
  \left.\frac{\Lambda}{S}\right|_{\tuSCR}=\frac{(3-\cos{2\shape^{\tuSCR}_{0}})\tan{\hat\shape}}{8\sin{\shape^{\tuSCR}_{0}}\sqrt{\tan^2\hat\shape-\tan^2\shape^{\tuSCR}_{0}}}.
\end{equation}
The tumbling solution only exists above a critical ratio $\left.\Lambda /
  S\right|_{\text{crit.}} \simeq 0.627$. This finding corroborates the first
order phase transition found numerically. An analytical expression for
$\left.\Lambda / S\right|_{\text{crit.}}$ can be derived by minimizing the
right hand side of 
(\ref{eq:lambdacrit}). However, the resulting
expression as a function of $\hat{\shape}$ is too involved to be reproduced
here.

The flow field of the shape parameter $\shape$ is shown in
figure~\ref{fig:cutshapea} for $S=10$. Below $\Lambda \simeq 6.27$, there is
only one (oscillating) stationary solution for $\shape$, which is the stable
transient motion. Above $\Lambda \simeq 6.27$, both the stable and unstable
tumbling branch and the swinging solution coexist. If we start with an initial
capsule shape $\shape(0)$ above the value $\shape^{\tuSCR}_{0}(\Lambda/S)$ on
the unstable branch, the capsule tumbles in the long time limit.  Starting
with an initial capsule shape $\shape(0)$ below the value
$\shape^{\tuSCR}_{0}(\Lambda/S)$ on the unstable branch, the capsule will
swing. To check the coexistence of both motions, we solved the dynamics now
starting with the initial shape $\shape(0)=\arcsin(1/\Lambda)$. With this
carefully chosen initial condition, transient motion is indeed stable for
arbitrarily large values of $\Lambda$.

\emph{Initial tumbling in the transient regime:}
Equations~(\ref{eq:tumblingeoma}--\ref{eq:tumblingeomc}) can also lead to more
insight regarding the transient regime. Since no tumbling solution is stable
for $\Lambda/S <0.627$, the shape parameter monotonously decreases until it
reaches $\shape_{0}=\arcsin 1/\Lambda\approx 1/\Lambda$. While $\shape(t)$ is
of the order one, the slow shape dynamics is given by
(\ref{eq:tumblingeoma}--\ref{eq:tumblingeomc}), which can be
easily solved numerically. It is, however, instructive to
consider the parameters $1\ll\Lambda\ll S$ well inside the
transient regime away from the phase boundaries. Here, the
equations of motion simplify to
\begin{eqnarray}
  \label{eq:transienteoma}
  \dot \rotangle^{\trSCR}_{0} &=& - \frac{1}{4\Lambda}
  \frac{1+\sin^{2}\shape_{0}}{\sin^{2}\shape_{0}},\\
  \label{eq:transienteomc}  \dot \shape^{\trSCR}_{0} &=& - \frac{1}{S} \cot\hat{\shape}
  \sin \shape_{0}.
\end{eqnarray}
We see that two slow time scales become relevant: The phase angle
$\rotangle$ changes monotonously on a typical time scale $\Lambda$,
while the shape relaxes on the much longer time scale $S$. During
this time the capsule tumbles. The decreasing phase angle
$\rotangle$ during the initial tumbling motion is also visible in
figure~\ref{fig:inclinationc}.

\section{Perspectives}
\label{sec:perspectives}

In this paper we have summarized the current state of research
concerning the dynamics of capsules in hydrodynamic shear flow.
We have reviewed and categorized the phenomenology observed in
experiments and in numerical simulations. In the review part, we
focused on the small deformation of initially spherical capsules
in shear flow and on the formation of wrinkles on polymerized
membranes. For initially non-spherical capsules, the shape memory
effect drastically alters the dynamics, leading to a swinging
motion. A reduced model with fixed shape explains the swinging
behaviour at low viscosity contrast, but is at variance with
numerical results at large shear rate. We have solved this
discrepancy by systematically expanding the equations of motion
for small deformations. Shape changes were found to be paramount
to understanding the dynamics. The disputed intermittent regime
was found to be an artefact of artificially fixing the capsule
shape. In the relevant parameter range the motion is rather a
transient one as suggested by numerical evidence. We have
obtained analytical expressions for the phase boundaries at strong
flow and have identified the swinging to transient transition as
being of second order, the transient to tumbling transition as
being of first order. In the tumbling regime a swinging motion is
also stable for carefully chosen initial conditions.

Until recently, all theoretical and experimental works have focused on soft
objects in time-constant flow. The dynamics within the reduced model with
constant shape was completely solved analytically for arbitrary time-dependent
flow in the quasi-spherical limit \cite{kessler2007}.  Here, the shear flow
oscillates around a finite positive value. With suitable modulation of the
shear rate, resonant tumbling motion can be excited, even though the mean
shear rate lies inside the swinging regime for constant flow. First
experiments in time-varying external flow considered red blood cells
\cite{takatani2006}, where the shear rate oscillates with a vanishing mean
value. The focus of this study was on the deformation and relaxation of the
shape. The oscillating shape showed a time lag with respect to the shear
stress of the flow. The shape changes were characterized by a rapid
deformation followed by a slow relaxation. Within the expanded reduced model,
Noguchi \cite{noguchi2009} found a qualitatively similar behaviour, deducing
that in the experiment the red blood cells have been in the low-frequency
swinging regime.  For vanishing mean shear rate, there is a rich phase
behaviour consisting of swinging and tumbling motions and shape changes. For
finite mean shear rates, the shape changes also lag behind the oscillating
shear stress. These results imply that shape changes might be important in
time-varying flow.

How the motion in time-dependent shear flow is altered by allowing shape
changes is an open question worthwhile pursuing.  We expect particularly
interesting effects in the tumbling regime, where two modes of motion are
simultaneously stable with their own basin of attraction. When changing the
flow parameters periodically, hysteresis might play an important role.

Shear flow above a substrate to which a capsule adheres could lead to
dynamical unbinding, or detachment. For vesicles, this effect has been studied
both theoretically \cite{cantat1999a, seifert1999b, sukumaran2001} and in
experiments \cite{lorz2000, abkarian2005a, chatkaew2009}. For capsules, the
equilibrium shape of such a bound configuration has been calculated as a
function of adhesion energy and material parameters \cite{graf2006,
  springman2008}. It would now be interesting to investigate how shear flow
affects such shapes and how shear-induced detachment of capsules differs from
that of vesicles.

A related, but different open area of research is the influence of thermal
fluctuations. While thermal noise is energetically not as relevant for
capsules as it is for vesicles, it is known to influence phase
transitions. Thermal fluctuations can be added to the equations of motion for
capsules as has been done for three-dimensional quasi-spherical vesicles
\cite{kats1996, haas1997, seifert1999a, noguchi2005a} and two-dimensional
quasi-circular vesicles \cite{finken2008, messlinger2009}. In Steinberg's
group~\cite{kantsler2005a}, thermally induced fluctuations in shape and
inclination angle of a giant vesicle in a stationary tank-treading regime was
investigated as a function of the membrane area and viscosity contrast.


Finally, it looks promising to treat the stationary motion of capsules and
vesicles as examples of non-equilibrium steady states from the perspective of
stochastic thermodynamics \cite{seifert2008, speck2008}. One could then ask
for the entropy production in such states and its behaviour at these dynamical
transitions. Likewise, the response of swinging and tumbling motion to
additional small perturbations could then be expressed by correlation
functions in the corresponding steady state \cite{seifert2010}. Within such a
perspective, vesicles and capsules could become two-dimensional paradigms for
exploring these novel concepts which so far have mainly been studied for
zero-dimensional and one-dimensional objects by using, respectively, colloidal
particles and polymers.

\appendix

\section{Vector spherical harmonics}
\label{sec:vect-spher-harm}
A complete orthonormal basis for three-dimensional vector fields on the sphere
is given by the vector spherical harmonics \cite{morse1953} which are defined
as
\begin{eqnarray}
  \vect Y^m_l(\theta,\phi) &\equiv& 
  Y^m_l(\theta,\phi) \, \vect e_r(\theta,\phi) \,, \\
  \vect \Psi^m_l(\theta,\phi) &\equiv& 
  \frac 1 {\sqrt{l(l+1)}} \hat\nabla Y^m_l(\theta,\phi) \,, \\
  \vect \Phi^m_l(\theta,\phi) &\equiv& 
  \vect e_r \times \vect \Psi^m_l(\theta,\phi) \,,
\end{eqnarray}
where $Y^m_l(\theta,\phi)$ are the standard scalar spherical harmonics
($l=0,1,\ldots$ and $m=-l,-l+1,\ldots,l$), the operator
\begin{eqnarray}
  \hat\nabla \equiv \vect e_\theta \partial_\theta 
  + \frac 1 {\sin\theta} \vect e_\phi \partial_\phi
\end{eqnarray}
is the surface gradient on the sphere, and the scalar product of complex vector
fields $\vect a$ and $\vect b$ is defined as
\begin{eqnarray}
  \langle \vect a, \vect b \rangle \equiv 
  \int_0^{2\pi}  \int_0^\pi 
  \vect a^\ast \cdot \vect b \,
  \sin\theta \, 
  \mathrm{d}\theta \,
  \mathrm{d}\phi
\end{eqnarray}
with index $\ast$ denoting complex conjugate.

\section{Elastic energy}
\label{sec:elastic-energy}
Each membrane point can be labelled by two Lagrangian coordinates
$(\thetaref,\phiref)$, or equivalently by points
$\Vx(\thetaref,\phiref)$ on a hypothetical reference sphere, if
$(\thetaref,\phiref)$ are interpreted as spherical coordinates.
Here,
\begin{equation}
  \label{eq:vxdef}
  \Vx(\thetaref,\phiref) \equiv R
  \transpose{[\sin(\thetaref)\cos(\phiref),\sin(\thetaref)\sin(\phiref),\cos(\thetaref)]}
\end{equation}
are the Cartesian coordinates of a point on the surface of a
sphere with radius $R$ with given spherical coordinates. The
actual membrane configuration is then constructed in two steps:
First, the reference sphere is rotated by a three-dimensional
rotation $\Vrot \equiv \rotanglelab \Vn$ by an angle $\rotanglelab$
around an axis of rotation $\Vn$. The position of the membrane
point is then given by $\rotmat_{\Vrot} \Vx(\thetaref,\phiref)
\equiv \Vx(\thetacoord,\phicoord)$. Here, $\rotmat_{\Vrot}$
represents the Cartesian rotation matrix. The analytical
expressions for the rotated spherical coordinates
$(\thetacoord,\phicoord)$ as a function of the original
coordinates $(\thetaref,\phiref)$ and the rotation angle $\Vrot$
are involved. Only in the special case where $\Vn=\Ve_{z}$ they
simplify to $\thetacoord=\thetaref$, $\phicoord=
\phiref+\rotanglelab$.

In a second step, the points of the rotated sphere are displaced
by a deformation vector $\epsilon R
\Vu(\thetacoord,\phicoord)$. The final position of a
membrane point is
\begin{equation}
    \Vr(\thetaref,\phiref)[\Vrot,\Vu] = \rotmat_{\Vrot}
    \Vx(\thetaref,\phiref) + \epsilon R \Vu(\thetacoord,\phicoord).
\end{equation}
We have factored the capsule radius out of the definition of
$\Vu$ to make $\Vu$ dimensionless. The pre-factor $\epsilon$
formally a small parameter for our perturbation expansion. Its
specific value will be chosen below for a convenient
normalisation of $\Vu$. The equilibrium shape of a capsule is
given by a specific deformation $\VU$ (we can set
$\VRot=\vect{0}$ without loss of generality).

The elastic energy of such a deformation is modelled using an
Evans-Skalak constitutive law with locally and globally
incompressible membrane \cite{pozrikidis2003book}. In the
quasi-spherical limit the energy can be written as
\begin{equation}
  \mathcal{H}[\Vu,\Vrot] \equiv \mu \int dA \,\Mat{\epsilon} :
  \Mat{\epsilon} + \int dA \sigma.
\end{equation}
The local tension $\sigma$ is a Lagrange multiplier to ensure
local and global area incompressibility. The strain tensor
\begin{equation}
  \Mat{\epsilon}(\thetaref,\phiref) \equiv \frac{1}{2}
  \left[\Mat{g}[\Vu](\theta,\phi) - \rotop_{\Vrot} \Mat{g}[\VU](\thetacoord,\phicoord) \right]
\end{equation}
measures the difference in the metric tensors $\Mat{g}$ between
the equilibrium and the actual configuration. Note that the
metrics are evaluated at the same material point
$(\thetaref,\phiref)$, which are in general at different
Cartesian positions in both configurations. One metric tensor
must therefore be moved to the other position. This is performed
using the rotation operator $\rotop_{\Vrot}$, which transforms
the tensor components accordingly. We can simplify the
calculation of the elastic energy considerably by taking
advantage of the fact that the strain tensor is linear in the
deformation in the quasi-spherical approximation. The strain
tensor and the corresponding elastic energy
$\mathcal{H}_{0}[\Vu]$ for a deformation from a \emph{spherical}
equilibrium shape ($\VU\equiv 0$) without area constraint has
already be calculated in the literature
\cite{barthes-biesel1980,rochal2005}. The corresponding
expression for a deformation from a non-spherical equilibrium
shape is then simply $\mathcal{H}_{0}[\Vu - \rotop_{\Vrot}
\VU]$.

To take advantage of the spherical symmetry of the reference sphere, we switch
to a vector spherical harmonics representation of the deformation field
\begin{equation}
  \label{eq:qssph}
  \Vu \equiv 
  \frac{u_{00}^Y}{\sqrt{4\pi}}\vect{e}_r + 
  \sum_{lm} \left[u^{Y}_{lm} \vect{Y}^{m}_{l} +
    u^{\Psi}_{lm} \vect{\Psi}^{m}_{l} + u^{\Phi}_{lm}
    \vect{\Phi}^{m}_{l} \right].
\end{equation}
Here and in the following, the sum over $l$ and $m$ runs from $l\geq 1$ with
$-l\leq m \leq l$. The vector spherical harmonics are defined in
\ref{sec:vect-spher-harm}. Expressions for the volume, area, and bending
energy in terms of the deformation amplitudes $u^{Y}_{lm}$ are derived in
\cite{seifert1999a}, while expressions for the elastic energy
against shearing and stretching in the case of spherical capsules can be found
in 
\cite{rochal2005}. 

%
To obey volume conservation, the constant $u_{00}$ is a function of the higher
mode amplitudes and can be eliminated.
The excess area is then given by 
\begin{equation}
  \Delta = \epsilon^2 \sum_{lm} |u^{Y}_{lm}|^{2} \frac{(l+2)(l-1)}{2} \equiv
  \frac{\epsilon^2}{2} \sum_{lm} F_l |u^{Y}_{lm}|^{2}.
\end{equation}
The shear modes $u^{\Phi}_{lm}$ completely decouple from the other two
deformation modes and are omitted in the following.  Taking advantage of the
local area constraint $2 (u^{Y}_{lm} - \rotop_{\Vrot} \U^{Y}_{lm}) = \sqrt{l(l+1)}
u^{\Psi}_{lm}$ we can write the elastic energy as
\begin{eqnarray}
  \mathcal{H}[\Vu,\Vrot] &=& \epsilon^{2} R^2 \frac{\mu}{2} \sum_{lm}
  \frac{4(l^2+l-2)}{l(l+1)} 
  |u^{Y}_{lm} - \rotop_{\Vrot} \U^{Y}_{lm}|^{2} 
  \\ \nonumber
  && 
  + \epsilon^{2} R^2 \frac{\bar{\sigma}}{2} \sum_{lm} F_{l} |u^{Y}_{lm}|^{2}.
\end{eqnarray}
Here, $\bar{\sigma}$ denotes the homogeneous tension $\bar{\sigma}
\equiv \int dA \sigma / A$ ensuring global area compressibility.

Although $\epsilon$ is formally a free parameter signifying the order
of magnitude of the terms, we chose to set
\begin{equation}
  \epsilon \equiv \sqrt{\Delta/2}.
\end{equation}
With this choice the normalization of the $\Vu_{lm}$ reads
\begin{equation}
  \frac{1}{4} \sum_{lm} F_{l} |u^{Y}_{lm}|^{2} = 1.
\end{equation}
When only $l=2$ modes are excited, this simplifies to $\sum_{m}
|u^{Y}_{2m}|^{2}=1$.

In the spherical harmonic representation the rotation operator
reduces to the Wigner rotation matrices
\begin{equation}
  \rotop_{\Vrot} u^{s}_{lm} = \sum_{m'} D^{l}_{m',m}(\Vrot) u^{s}_{lm'}.
\end{equation}
The elastic force in radial direction is given by
\begin{eqnarray}
\label{eq:feq}
  {f}^{Y\el}_{lm} &\equiv &
  - \frac{\partial H[\Vu]}{
    \epsilon R^3 \partial {u}^{Y\ast}_{lm}} \\
  &=& - \frac{\mu \epsilon}{R}
  \frac{4(l^2+l-2)}{l(l+1)} 
  \left[ u^{Y}_{lm} - \sum D^{l}_{mm'}(\Vrot) {\U}_{lm'}\right]
  - \frac{\epsilon \bar{\sigma}}{R} F_l u^{Y}_{lm}.
  \nonumber
\end{eqnarray}
where the global area constraint contributes a tensile force.
Additional
forces ensure local area incompressibility under the capsule motion.

\section{Equations of motion}
\label{sec:equations-motion}

Since Stokes' equations are linear, any solution can be written
as the superposition of different flow fields. Moreover, we can
chose the boundary conditions at the membrane and at infinity of
each particular solution to our advantage, as long as the overall
solution matches the correct boundary conditions given by the
force balance and the external flow. We can thus separate the
flow fields induced by an elongational and rotational part and
flow induced by the membrane forces and torques
\begin{equation}
  \Vv(\Vr) = \Vv_{\text{ind}}(\Vr) + \Vv_{s}(\Vr) + \Vv_{\omega}(\Vr).
\end{equation}

For the rotational velocity field $\Vv_{\omega}$ we take the corresponding
component of the external flow throughout
\begin{equation}
  \vect{v}_{\omega} = \omega \vect{e}_z \times \vect{r}.
\end{equation}
Since the hydrodynamic stress of a purely rotating fluid
vanishes, this flow does not induce forces at the membrane. As
the elastic energy of the membrane depends on $\Vrot$, a torque
is exerted on the surrounding fluid. This torque is, however, of
the order $\epsilon^{2}$ and is neglected in the following. The
no-slip boundary condition at the capsule membrane therefore
requires
\begin{equation}
  \label{eq:nosliprot}
  \dot{\Vrot} = \omega \Ve_{z}.
\end{equation}
For $\Vv_{s}$ we chose a flow which matches the elongational part of the
external flow at infinity and vanishes at the sphere.  While such a flow field
produces surface forces on the capsule membrane, these forces only depend on
the outer viscosity $\etaout$. The resulting force at $r=R$ is
\begin{equation}
  \label{eq:fs}
  \Vf^s_{2,m} = 5 i s \etaout  (\delta_{m,-2}-\delta_{m,2})
  \sqrt{\frac{2\pi}{15}} \left(2 \vect{Y}^m_2 +\sqrt{6}\vect{\Psi}^m_2\right).
\end{equation}

The remaining flow field is induced by the surface forces of
the capsule membrane and of the hydrodynamic stress jump of
the extensional flow. The induced flow field itself vanishes
away from the capsule. In other words, the induced flow field
is the same as in an ambient fluid. The global flow field can
be found explicitly (see \cite{seifert1999a}), but is of no
interest here. We only need to know the value of the flow
field at the capsule. The linear relationship between the
surface forces and the induced velocity is given by the (inverse) Oseen
tensor
\begin{equation}
  \label{eq:vind}
  \Oseen^{-1}_l \cdot \Vv^{\text{ind}}_{l,m} =
  \Vf^s_{l,m} + \Vf^{\el}_{l,m} + \Vf^{\sigma}_{l,m}.
\end{equation}
The viscosity contrast only appears in the expression of the
Oseen tensor, which reads explicitly \cite{rochal2005}
\begin{equation}
  \Oseen^{-1}_l \equiv \frac{1}{R} \left(\begin{array}{cc}
      \etain \frac{2l^2+l+3}{l} + \etaout \frac{2l^2+3l+4}{l+1} &
      -3 \etain \sqrt{\frac{l+1}{l}} - 3 \etaout
      \sqrt{\frac{l}{l+1}}\\
      -3 \etain \sqrt{\frac{l+1}{l}} - 3 \etaout  \sqrt{\frac{l}{l+1}}& \etain (2l+1) +
      \etaout (2l+1)
    \end{array}\right)
  .
\end{equation}
Again, local area incompressibility enforces a coupling between
radial and tangential flow.

Finally, the membrane is advected with the resulting flow owing to the no-slip
boundary condition. 

The membrane dynamics is thus governed by the equation 
\begin{equation}
  D_t \vect u_{lm}
  =
  \Vv^{s}_{lm}
  + \Oseen_l
  \left( 
    \Vf^s_{l,m} + \Vf^{\el}_{l,m} + \Vf^{\sigma}_{l,m}
  \right) .
  \label{eq:eom}
\end{equation}
Here, the material time derivative of the deformation amplitude
reads
\begin{equation}
  D_t \Vu \equiv (\partial_t + (\Vv \cdot \vect{\nabla})) \Vu.
\end{equation}
In the following we only take the rotational part of the flow
when calculating the advection, since the other contributions are smaller by a
factor $\epsilon$.  In other words, we use
\begin{equation}
  \label{eq:advection}
  D_t \Vu = (\partial_t + (\dot{\Vrot} \times \vect{r})
  \cdot \vect{\nabla}) \Vu = (\partial_t + \omega
  \partial_{\phi}) \Vu.
\end{equation}
This is justified because the rotational velocity is an order of magnitude
larger than the velocity components leading to a membrane deformation.
Equations~(\ref{eq:feq}, \ref{eq:nosliprot}, \ref{eq:fs}, \ref{eq:vind},
\ref{eq:eom}, \ref{eq:advection})
specify the dynamics of the capsule in the quasi-spherical limit completely.

We see that in linear external flow only the $l=2$ modes can be excited.
We therefore restrict ourselves to only taking $l=2$ deformation modes. Since
the radial and tangential modes are coupled by the local area constraint, we
can eliminate the latter modes. By redefining the time scale $\tau$ and
introducing dimensionless strengths of the rotational $\Lambda$ and
elongational $S$ flow analogously to 
\cite{lebedev2008} (see
(\ref{eq:paramsevans})) the equations of motions simplify to (omitting the now
redundant $Y$ and $l=2$ indices) 
\begin{eqnarray}
  \partial_{\tau}\phi &=& \Lambda \\
  (\partial_\tau - i \Lambda m) u_m &=& \sum_{m'}
  \left(\delta_{mm'} - u_{m} u_{m'}^\ast\right) \times
  \\ \nonumber 
  &&
  \left[
    S^{-1} \exp(-i m'\rotanglelab) \U_{m'}/\sin{\hat\shape}
    - i (\delta_{m',2}-\delta_{m',-2})
  \right]
  .
\end{eqnarray}
These equations are formally very similar to the equations of
motions for vesicles in 
(48) of 
\cite{lebedev2008}.
Let us first search for solutions which show the same mirror
symmetry with respect to the $xy$-plane as the external flow.
Thus, only the modes with $m=-2,0,2$ are non-zero (also for
$\U_{m}$). 
Since the modes are normalized to $\sum_{m} |u_{m}|^{2}=1$, we
can employ the parametrisation
\begin{eqnarray}
  u_{0} &\equiv& \cos(\shape)\\
  u_{2} &\equiv& \sin(\shape) \exp(- 2 i \inclSK) / \sqrt{2}\\
  u_{-2} &\equiv& \sin(\shape) \exp(2 i \inclSK) / \sqrt{2}
\end{eqnarray}
Here, $\inclSK \in [-\pi/2,\pi/2]$ is the inclination angle of
the capsule with respect to the $x$-axis, $\shape \in [0,\pi/2]$.
To compare with the equations of motion
(\ref{eq:skotheimeoma}, \ref{eq:skotheimeomb}) in the reduced
Skotheim model, (cf. section \ref{sec:reducedmodel}) it is
advantageous to switch to the phase angle $\rotangle$ in the
capsule frame rather than the phase angle $\rotanglelab$ in the
laboratory frame. Both angles are related by the inclination
angle, $\rotanglelab\equiv\rotangle+\inclSK$. We thus arrive at
the equations of motion (\ref{eq:qseoma}--\ref{eq:qseomc}).



\section*{References}

\begin{thebibliography}{10}

\bibitem{pozrikidis2003book}
C.~Pozrikidis, editor.
\newblock {\em Modelling and simulation of capsules and biolocical cells}.
\newblock Chapman \& Hall/CRC, 2003.

\bibitem{moehwald2003}
H.~M{\"o}hwald, E.~Donath, and G.~Sukhorukov.
\newblock {\em Smart capsules}.
\newblock New York: Wiley VCH, 2003.

\bibitem{fery2007}
A.~Fery and R.~Weinkamer.
\newblock Mechanical properties of micro- and nanocapsules: Single-capsule
  measurements.
\newblock {\em Polymer}, 48:7221, 2007.

\bibitem{abkarian2008}
M.~Abkarian and A.~Viallat.
\newblock Vesicles and red blood cells in shear flow.
\newblock {\em Soft matter}, 4:653, 2008.

\bibitem{barthes-biesel2009}
D.~Barth{\`e}s-Biesel.
\newblock Capsule motion in flow: Deformation and membrane buckling.
\newblock {\em C. R. Physique}, 10:764, 2009.

\bibitem{seifert1997}
U.~Seifert.
\newblock Configurations of fluid membranes and vesicles.
\newblock {\em Adv. Physics}, 46:13, 1997.

\bibitem{walter2002}
H.~Rehage, M.~Husmann, and A.~Walter.
\newblock From two-dimensional model networks to microcapsules.
\newblock {\em Rheol. Acta}, 41:292, 2002.

\bibitem{Discher2001}
D.~E. Discher and P.~Carl.
\newblock New insights into red cell network structure, elasticity, and
  spectrin unfolding - a current review.
\newblock {\em Cell. Mol. Biol. Lett.}, 6:593, 2001.

\bibitem{patwardhan1983}
S.~A. Patwardhan and K.~G. Das.
\newblock Microencapsulation.
\newblock In K.~G. Das, editor, {\em Controlled release technology -
  bioengineering aspects}, pages 121--141. Wiley, New York, 1983.

\bibitem{dinsmore2002}
A.~D. Dinsmore, M.~F. Hsu, M.~G. Nikolaides, M.~Marquez, A.~R. Bausch, and D.A.
  Weitz.
\newblock Colloidosomes: Selectively permeable capsules composed of colloidal
  particles.
\newblock {\em Science}, 298:1006, 2002.

\bibitem{keller1982}
S.~R. Keller and R.~Skalak.
\newblock Motion of a tank-treading ellipsoidal particle in a shear-flow.
\newblock {\em J. Fluid. Mech.}, 120:27, 1982.

\bibitem{fischer2004}
T.~M. Fischer.
\newblock Shape memory of human red blood cells.
\newblock {\em Biophys. J.}, 86:3304, 2004.

\bibitem{skotheim2007}
J.~M. Skotheim and T.~W. Secomb.
\newblock Red blood cells and other nonspherical capsules in shear flow:
  Oscillatory dynamics and the tank-treading-to-tumbling transition.
\newblock {\em Phys. Rev. Lett.}, 98:078301, 2007.

\bibitem{abkarian2007}
M.~Abkarian, M.~Faivre, and A.~Viallat.
\newblock Swinging of red blood cells under shear flow.
\newblock {\em Phys. Rev. Lett.}, 98:188302, 2007.

\bibitem{rochal2005}
S.~B. Rochal, V.~L. Lorman, and G.~Mennessier.
\newblock Viscoelastic dynamics of spherical composite vesicles.
\newblock {\em Phys. Rev. E}, 71, 2005.

\bibitem{barthes-biesel1980}
D.~Barth{\`e}s-Biesel.
\newblock Motion of a spherical microcapsule freely suspended in a linear shear
  flow.
\newblock {\em J. Fluid. Mech.}, 100:831, 1980.

\bibitem{barthes-biesel1981}
D.~Barth{\`e}s-Biesel and J.~M. Rallison.
\newblock The time-dependent deformation of a capsule freely suspended in a
  linear shear flow.
\newblock {\em J. Fluid. Mech.}, 113:251, 1981.

\bibitem{takatani2006}
N.~Watanabe, H.~Kataoka, T.~Yasuda, and S.~Takatani.
\newblock Dynamic deformation and recovery response of red blood cells to a
  cyclically reversing shear flow: Effects of frequency of cyclically reversing
  shear flow and shear stress level.
\newblock {\em Biophys. J.}, 91:1984, 2006.

\bibitem{fischer2007}
T.~M. Fischer.
\newblock Tank-tread frequency of the red cell membrane: Dependence on the
  viscosity of the suspending medium.
\newblock {\em Biophys. J.}, 93:2553, 2007.

\bibitem{misbah2006}
C.~Misbah.
\newblock Vacillating breathing and tumbling of vesicles under shear flow.
\newblock {\em Phys. Rev. Lett.}, 96:28104, 2006.

\bibitem{noguchi2007}
H.~Noguchi and G.~Gompper.
\newblock Swinging and tumbling of fluid vesicles in shear flow.
\newblock {\em Phys. Rev. Lett.}, 98:128103, 2007.

\bibitem{lebedev2007a}
V.~V. Lebedev, K.~S. Turitsyn, and S.~S. Vergeles.
\newblock Dynamics of nearly spherical vesicles in an external flow.
\newblock {\em Phys. Rev. Lett.}, 99:218101, 2007.

\bibitem{li1988}
X.~Z. Li, D.~Barth{\`e}s-Biesel, and A.~Helmy.
\newblock Large deformations and burst of a capsule freely suspended in an
  elongational flow.
\newblock {\em J. Fluid. Mech.}, 187:179, 1988.

\bibitem{leyrat-maurin1993}
A.~Leyrat-Maurin, A.~Drochon, and D.~Barthes-Biesel.
\newblock Flow of a capsule through a constriction - application to cell
  filtration.
\newblock {\em J. Phys. Paris III}, 3:1051, 1993.

\bibitem{leyrat-maurin1994}
A.~Leyrat-Maurin and D.~Barthes-Biesel.
\newblock Motion of a deformable capsule through a hyperbolic constriction.
\newblock {\em J. Fluid. Mech.}, 279:135, 1994.

\bibitem{zhou95}
H.~Zhou and C.~Pozrikidis.
\newblock Deformation of liquid capsules with incompressible interfaces in
  simple shear flow.
\newblock {\em J. Fluid. Mech.}, 283:175, 1995.

\bibitem{pozrikidis1995}
C.~Pozrikidis.
\newblock Finite deformation of liquid capsules enclosed by elastic membranes
  in simple shear-flow.
\newblock {\em J. Fluid. Mech.}, 297:123, 1995.

\bibitem{ramanujan1998}
S.~Ramanujan and C.~Pozrikidis.
\newblock Deformation of liquid capsules enclosed by elastic membranes in
  simple shear flow: Large deformations and the effect of fluid viscosities.
\newblock {\em J. Fluid Mech.}, 361:117, 1998.

\bibitem{pozr01}
C.~Pozrikidis.
\newblock Effect of bending stiffness on the deformation of liquid capsules
  enclosed by elastic membranes in simple shear flow.
\newblock {\em J. Fluid. Mech.}, 440:269, 2001.

\bibitem{pozrikidis2003a}
C.~Pozrikidis.
\newblock Numerical simulation of the flow-induced deformation of red blood
  cells.
\newblock {\em Ann. Biomed. Eng.}, 31:1194, 2003.

\bibitem{lac03}
E.~Lac.
\newblock {\em D\'eformation et convection d'une capsule dans un \'ecoulement
  de {S}tokes tridimenionnel infini}.
\newblock PhD thesis, Universit\'e Compiegne, 2003.

\bibitem{lac2004}
E.~Lac, D.~Barthès-Biesel, N.~A. Pelekasis, and J.~Tsamopoulos.
\newblock Spherical capsules in three-dimensional unbounded {S}tokes flows:
  effect of the membrane constitutive law and onset of buckling.
\newblock {\em J. Fluid. Mech.}, 516:303, 2004.

\bibitem{lac2005}
E.~Lac and D.~Barth{\`e}s-Biesel.
\newblock Deformation of a capsule in shear flow: Effect of membrane prestress.
\newblock {\em Phys. Fluids}, 17:72105, 2005.

\bibitem{kessler2007}
S.~Kessler, R.~Finken, and U.~Seifert.
\newblock Swinging and tumbling of elastic capsules in shear flow.
\newblock {\em J. Fluid. Mech.}, 605:207, 2007.

\bibitem{pozrikidis2006a}
C.~Pozrikidis.
\newblock A spectral collocation method with triangular boundary elements.
\newblock {\em Eng. Anal. Bound. Elem.}, 30:315, 2006.

\bibitem{Wang2006}
Y.~C. Wang and P.~Dimitrkopoulos.
\newblock {A three-dimensional spectral boundary element algorithm for
  interfacial dynamics in {S}tokes flow}.
\newblock {\em Phys. Fluids}, 18:82106, 2006.

\bibitem{dimitrakopoulos2007}
P.~Dimitrakopoulos.
\newblock Interfacial dynamics in {S}tokes flow via a three-dimensional
  fully-implicit interfacial spectral boundary element algorithm.
\newblock {\em J. Comput. Phys.}, 225:408, 2007.

\bibitem{dodson2008}
W.~R. Dodson~III and P.~Dimitrakopoulos.
\newblock Spindles, cusps, and bifurcation for capsules in {S}tokes flow.
\newblock {\em Phys. Rev. Lett.}, 101:208102, 2008.

\bibitem{peskin1977}
C.~S. Peskin.
\newblock Numerical-analysis of blood-flow in heart.
\newblock {\em J. Comput. Phys.}, 25:220, 1977.

\bibitem{peskin2002}
C.~S. Peskin.
\newblock The immersed boundary method.
\newblock {\em Acta Numerica}, page 479, 2002.

\bibitem{eggleton1998}
C.~D. Eggleton and A.~S. Popel.
\newblock Large deformation of red blood cell ghosts in a simple shear flow.
\newblock {\em Phys. Fluids}, 10:1834, 1998.

\bibitem{bagchi2009}
P.~Bagchi and R.~M. Kalluri.
\newblock {Dynamics of nonspherical capsules in shear flow}.
\newblock {\em Phys. Rev. E}, {80}:{016307}, {2009}.

\bibitem{sui2007}
Y.~Sui, Y.~T. Chew, P.~Roy, X.~B. Chen, and H.~T. Low.
\newblock Transient deformation of elastic capsules in shear flow: Effect of
  membrane bending stiffness.
\newblock {\em Phys. Rev. E}, 75:066301, 2007.

\bibitem{sui2007a}
Y.~Sui, Y.~T. Chew, and H.~T. Low.
\newblock {A lattice Boltzmann study on the large deformation of red blood
  cells in shear flow}.
\newblock {\em {Int. J. Mod. Phys. C}}, {18}:{993}, {2007}.

\bibitem{sui2007b}
Y.~Sui, Y.~T. Chew, P.~Roy, and H.~T. Low.
\newblock {Effect of membrane bending stiffness on the deformation of elastic
  capsules in extensional flow: A lattice Boltzmann study}.
\newblock {\em Int. J. Mod. Phys. C}, {18}:{1277}, {2007}.

\bibitem{sui2008}
Y.~Sui, H.~T. Low, Y.~T. Chew, and P.~Roy.
\newblock Tank-treading, swinging, and tumbling of liquid-filled elastic
  capsules in shear flow.
\newblock {\em Phys. Rev. E}, 77:016310, 2008.

\bibitem{sui2008a}
Y.~Sui, Y.~T. Chew, P.~Roy, Y.~P. Cheng, and H.~T. Low.
\newblock Dynamic motion of red blood cells in simple shear flow.
\newblock {\em Phys. Fluids}, 20:112106, 2008.

\bibitem{sui2010}
Y.~Sui, X.~B. Chen, Y.~T. Chew, P.~Roy, and H.~T. Low.
\newblock Numerical simulation of capsule deformation in simple shear flow.
\newblock {\em Comput. Fluids}, 39:242, 2010.

\bibitem{sui2010a}
Y.~Sui, H.T. Low, Y.~T. Chew, and P.~Roy.
\newblock A front-tracking lattice boltzmann method to study flow-induced
  deformation of three-dimensional capsules.
\newblock {\em Computers \& Fluids}, 39:499, 2010.

\bibitem{unverdi1992}
S.~U. Unverdi and G.~Tryggvason.
\newblock A front-tracking method for viscous, incompressible, multi-fluid
  flows.
\newblock {\em J. Comput. Phys.}, 100:25, 1992.

\bibitem{ma2009}
G.~Ma, J.~S. Hua, and H.~Li.
\newblock Numerical modeling of the behavior of an elastic capsule in a
  microchannel flow: The initial motion.
\newblock {\em Phys. Rev. E}, 79:046710, 2009.

\bibitem{low2007}
Y.~Sui, Y.~T. Chew, P.~Roy, X.~B. Chen, and H.~T. Low.
\newblock Transient deformation of elastic capsules in shear flow: Effect of
  membrane bending stiffness.
\newblock {\em Phys. Rev. E}, 75:66301, 2007.

\bibitem{kaoui2008}
B.~Kaoui, G.~H. Ristow, I.~Cantat, C.~Misbah, and W.~Zimmermann.
\newblock Lateral migration of a two-dimensional vesicle in unbounded
  poiseuille flow.
\newblock {\em Phys. Rev. E}, 77:021903, 2008.

\bibitem{kraus1996}
M.~Kraus, W.~Wintz, U.~Seifert, and R.~Lipowsky.
\newblock Fluid vesicles in shear flow.
\newblock {\em Phys. Rev. Lett.}, 77:3685, 1996.

\bibitem{biben2009}
T.~Biben, A.~Farutin, and C.~Misbah.
\newblock Numerical study of 3d vesicles under flow: discovery of new peculiar
  behaviors.
\newblock {\em arXiv:0912.4702 [cond-mat.soft]}, 2009.

\bibitem{biben2003a}
T.~Biben and C.~Misbah.
\newblock Tumbling of vesicles under shear flow within an advected-field
  approach.
\newblock {\em Phys. Rev. E}, 67:031908, 2003.

\bibitem{beaucourt2004b}
J.~Beaucourt, F.~Rioual, T.~Seon, T.~Biben, and C.~Misbah.
\newblock Steady to unsteady dynamics of a vesicle in a flow.
\newblock {\em Phys. Rev. E}, 69:011906, 2004.

\bibitem{biben2005}
T.~Biben, K.~Kassner, and C.~Misbah.
\newblock Phase-field approach to three-dimensional vesicle dynamics.
\newblock {\em Phys. Rev. E}, 72:041921, 2005.

\bibitem{noguchi2004}
H.~Noguchi and G.~Gompper.
\newblock Fluid vesicles with viscous membranes in shear flow.
\newblock {\em Phys. Rev. Lett.}, 93:258102, 2004.

\bibitem{noguchi2005}
H.~Noguchi and G.~Gompper.
\newblock Shape transitions of fluid vesicles and red blood cells in capillary
  flows.
\newblock {\em Proc. Natl. Acad. Sci. USA}, 102:14159, 2005.

\bibitem{noguchi2005a}
H.~Noguchi and G.~Gompper.
\newblock Dynamics of fluid vesicles in shear flow: Effect of membrane
  viscosity and thermal fluctuations.
\newblock {\em Phys. Rev. E}, 72:011901, 2005.

\bibitem{mcwhirter2009}
J.~L. McWhirter, H.~Noguchi, and G.~Gompper.
\newblock Flow-induced clustering and alignment of vesicles and red blood cells
  in microcapillaries.
\newblock {\em Proc. Natl. Acad. Sci. USA}, 106:6039, 2009.

\bibitem{chang1993b}
K.~S. Chang and W.~L. Olbricht.
\newblock Experimental studies of the deformation of a synthetic capsule in
  extensional flow.
\newblock {\em J. Fluid. Mech.}, 250:587, 1993.

\bibitem{chang1993}
K.~S. Chang and W.~L. Olbricht.
\newblock Experimental studies of the deformation and breakup of a synthetic
  capsule in steady and unsteady simple shear-flow.
\newblock {\em J. Fluid. Mech.}, 250:609, 1993.

\bibitem{walter2000}
A.~Walter, H.~Rehage, and H.~Leonhard.
\newblock Shear-induced deformations of polyamid microcapsules.
\newblock {\em Colloid. Polym. Sci.}, 278:169, 2000.

\bibitem{walter2001}
A.~Walter, H.~Rehage, and H.~Leonhard.
\newblock Shear induced deformation of microcapsules: shape oscillations and
  membrane folding.
\newblock {\em Coll. Surf. A}, 183-185:123, 2001.

\bibitem{goldsmith1972}
H.~L. Goldsmith and J~Marlow.
\newblock Flow behavior of erythrocytes.~1.~rotation and deformation in dilute
  suspensions.
\newblock {\em Proc. R. Soc. London, Ser. B}, 182:351, 1972.

\bibitem{fischer1977}
T.~Fischer and H.~Schmid-Sch\"onbein.
\newblock Tank tread motion of red-cell membranes in viscometric flow -
  behavior of intracellular and extracellular markers (with film).
\newblock {\em Blood Cells}, 3:351, 1977.

\bibitem{fischer1978}
T.~Fischer, M.~St\"ohrliesen, and H.~Schmid-Sch\"onbein.
\newblock Red-cell as a fluid droplet - tank tread-like motion of human
  erythrocyte-membrane in shear-flow.
\newblock {\em Science}, 202:894, 1978.

\bibitem{transontay1984}
R.~Tran-Son-Tay, S.~P. Sutera, and P.~R. Rao.
\newblock Determination of red-blood-cell membrane viscosity from rheoscopic
  observations of tank-treading motion.
\newblock {\em Biophys. J.}, 46:65, 1984.

\bibitem{cerda2003}
E.~Cerda and L.~Mahadevan.
\newblock Geometry and physics of wrinkling.
\newblock {\em Phys. Rev. Lett.}, 90:074302, 2003.

\bibitem{finken2006}
R.~Finken and U.~Seifert.
\newblock Wrinkling of microcapsules in shear flow.
\newblock {\em J. Phys.: Condens. Matter}, 18:L185, 2006.

\bibitem{kessler2009}
S.~Kessler, R.~Finken, and U.~Seifert.
\newblock Elastic capsules in shear flow: Analytical solutions for constant and
  time-dependent shear rates.
\newblock {\em Eur. Phys. J. E}, 29:399, 2009.

\bibitem{noguchi2009a}
H.~Noguchi.
\newblock Swinging and synchronized rotations of red blood cells in simple
  shear flow.
\newblock {\em Phys. Rev. E}, 80:021902, 2009.

\bibitem{hinch1991}
E.~J. Hinch.
\newblock {\em Perturbation Methods}, chapter~7.
\newblock Cambridge University Press, 1991.

\bibitem{noguchi2010b}
H.~Noguchi.
\newblock Dynamic modes of microcapsules in steady shear flow: Effects of
  bending and shear elasticities.
\newblock {\em arXiv:1003.3152v1 [cond-mat.soft]}, 2010.

\bibitem{deschamps2009b}
J.~Deschamps, V.~Kantsler, E.~Segre, and V.~Steinberg.
\newblock Dynamics of a vesicle in general flow.
\newblock {\em Proc. Natl. Acad. Sci. USA}, 106:11444, 2009.

\bibitem{noguchi2009}
H.~Noguchi.
\newblock Dynamic modes of red blood cells in steady and oscillatory shear
  flows.
\newblock {\em arXiv:0903.0038v1 [cond-mat.soft]}, 2009.

\bibitem{cantat1999a}
I.~Cantat and C.~Misbah.
\newblock Lift force and dynamical unbinding of adhering vesicles under shear
  flow.
\newblock {\em Phys. Rev. Lett.}, 83:880, 1999.

\bibitem{seifert1999b}
U.~Seifert.
\newblock Hydrodynamic lift on bound vesicles.
\newblock {\em Phys. Rev. Lett.}, 83:876, 1999.

\bibitem{sukumaran2001}
S.~Sukumaran and U.~Seifert.
\newblock Influence of shear flow on vesicles near a wall: A numerical study.
\newblock {\em Phy. Rev. E}, 64:11916, 2001.

\bibitem{lorz2000}
B.~Lorz, R.~Simson, J.~Nardi, and E.~Sackmann.
\newblock Weakly adhering vesicles in shear flow: Tanktreading and anomalous
  lift force.
\newblock {\em Europhys. Lett.}, 51:468, 2000.

\bibitem{abkarian2005a}
M.~Abkarian and A.~Viallat.
\newblock Dynamics of vesicles in a wall-bounded shear flow.
\newblock {\em Biophys. J.}, 89:1055, 2005.

\bibitem{chatkaew2009}
S.~Chatkaew, M.~Georgelin, M.~Jaeger, and M.~Leonetti.
\newblock Dynamics of vesicle unbinding under axisymmetric flow.
\newblock {\em Phys. Rev. Lett.}, 103:248103, 2009.

\bibitem{graf2006}
P.~Graf, R.~Finken, and U.~Seifert.
\newblock Adhesion of microcapsules.
\newblock {\em Langmuir}, 22:7117, 2006.

\bibitem{springman2008}
R.~M. Springman and J.~L. Bassani.
\newblock Snap transitions in adhesion.
\newblock {\em J. Mech. Phys. Solids}, 56:2358, 2008.

\bibitem{kats1996}
E.~I. Kats, V.~V. Lebedev, and A.~R. Muratov.
\newblock Nearly spherical vesicles: Shape fluctuations.
\newblock {\em JETP Lett.}, 63:216, 1996.

\bibitem{haas1997}
K.~H. de~Haas, C.~Blom, D.~van~den Ende, M.~H.~G. Duits, and J.~Mellema.
\newblock Deformation of giant lipid bilayer vesicles in shear flow.
\newblock {\em Phys. Rev. E}, 56:7132, 1997.

\bibitem{seifert1999a}
U.~Seifert.
\newblock Fluid dynamics in hydrodynamic force fields: Formalism and an
  application to fluctuating quasispherical vesicles in shear flow.
\newblock {\em Eur. Phys. J. B}, 8:405, 1999.

\bibitem{finken2008}
R.~Finken, A.~Lamura, U.~Seifert, and G.~Gompper.
\newblock Two-dimensional fluctuating vesicles in linear shear flow.
\newblock {\em Eur. Phys. J. E}, 25:309, 2008.

\bibitem{messlinger2009}
S.~Messlinger, B.~Schmidt, H.~Noguchi, and G.~Gompper.
\newblock {Dynamical regimes and hydrodynamic lift of viscous vesicles under
  shear}.
\newblock {\em Phys. Rev. E}, {80}:{011901}, {2009}.

\bibitem{kantsler2005a}
V.~Kantsler and V.~Steinberg.
\newblock Orientation and dynamics of a vesicle in tank-treading motion in
  shear flow.
\newblock {\em Phys. Rev. Lett.}, 95:258101, 2005.

\bibitem{seifert2008}
U.~Seifert.
\newblock Stochastic thermodynamics: principles and perspectives.
\newblock {\em Eur. Phys. J. B}, 64:423, 2008.

\bibitem{speck2008}
T.~Speck, J.~Mehl, and U.~Seifert.
\newblock Role of external flow and frame invariance in stochastic
  thermodynamics.
\newblock {\em Phys. Rev. Lett.}, 100:178302, 2008.

\bibitem{seifert2010}
U.~Seifert and T.~Speck.
\newblock Fluctuation-dissipation theorem in nonequilibrium steady states.
\newblock {\em Europhys. Lett}, 89:10007, 2010.

\bibitem{morse1953}
P.~M. Morse and H.~Feshbach.
\newblock {\em Methods of Theoretical Physics}, volume Part II.
\newblock McGraw-Hill, 1953.
\newblock p. 1298.

\bibitem{lebedev2008}
V.~Lebedev, K.~Turitsyn, and S.~Vergeles.
\newblock Nearly spherical vesicles in an external flow.
\newblock {\em New Journal of Physics}, 10:043044, 2008.

\end{thebibliography}
\end{document}